%% file: Draft-v2.tex
\documentclass[aps, prl, 10pt, twocolumn, superscriptaddress, noshowpacs, preprintnumbers, floatfix]{revtex4-1}

\usepackage[colorlinks=true,breaklinks=true]{hyperref}
\hypersetup{allcolors=[rgb]{0.0 0.0 1.0},linkcolor=[rgb]{0.75 0.05 0.05}}
\usepackage[dvipsnames]{xcolor}
\usepackage{mathtools}
\usepackage{amsmath}
\usepackage{amsfonts}
\usepackage{amssymb}
\usepackage{bbold}
\usepackage{multirow}
\usepackage{bm}
\long\def\exclude#1{}
\usepackage{orcidlink}

\newcommand{\bD}{{\bf D}}

\newcommand{\bB}{{\bf B}}
\newcommand{\bP}{{\bf P}}

\newcommand{\bp}{{\bf p}}
\newcommand{\bj}{{\bf j}}
\newcommand{\br}{{\bf r}}

\newcommand{\bq}{{\bf q}}

\newcommand{\bK}{{\bf K}}

\newcommand{\be}{{\bf e}}

\newcommand{\bv}{{\bf v}}

\newcommand{\bE}{{\bf E}}

\newcommand{\GF}{G_{\rm F}}

\begin{document}

\title{Collective flavor conversions are interactions of neutrinos with quantized flavor waves}

\author{Damiano F.\ G.\ Fiorillo \orcidlink{0000-0003-4927-9850}}
\affiliation{Deutsches Elektronen-Synchrotron DESY,
Platanenallee 6, 15738 Zeuthen, Germany}

\author{Georg G.\ Raffelt
\orcidlink{0000-0002-0199-9560}}
\affiliation{Max-Planck-Institut f\"ur Physik, Boltzmannstr.~8, 85748 Garching, Germany}

\begin{abstract}
Collective oscillations in dense neutrino gases (flavor waves) are notable for their instabilities that cause fast flavor conversion. We develop a quantum theory of interacting neutrinos and flavor wave quanta, which are analogous to plasmons, but also carry flavor. The emission or absorption of such flavor plasmons $\psi$, or \textit{flavomons}, changes the neutrino flavor. When an angular crossing occurs, the process $\nu_\mu\to\nu_e+\psi$ is more rapid than its inverse along the direction of the crossing, triggering stimulated $\psi$ emission and fast instability. Calculating the rate via Feynman diagrams matches the fast instability growth rate. Our novel $\nu$ and $\psi$ kinetic equations, corresponding to quasi-linear theory, describe instability evolution without resolving the small scales of the flavomon wavelength, potentially overcoming the main challenge of fast flavor evolution.
\end{abstract}


\maketitle

{\bf\textit{Introduction.}}---A system of many interacting particles can show collective behavior even when collisions are rare and the gas is not a fluid. In astrophysics, such collisionless plasmas abound, where electrons and ions interact through the coherent electric field they source. The modes of collective oscillations thus supported are elementary excitations on their own, but exist only in the medium. The simplest example is the longitudinal plasmon, the quantum of electrostatic oscillations first introduced by Pines in 1956~\cite{pines1956collective}.

Intriguingly, a similar picture emerges for a collisionless gas of neutrinos despite their very feeble \hbox{interactions}. Neutrino-neutrino refraction, once more without collisions, spawns collective forms of flavor evolution \hbox{\cite{Pantaleone:1992eq, Samuel:1993uw, Samuel:1995ri, Duan:2006an, Sawyer:2004ai, Izaguirre:2016gsx, Mirizzi:2015eza, Tamborra:2020cul, Richers:2022zug, Volpe:2023met}}, notably in core-collapse supernovae and neutron-star mergers, where refractive flavor exchange can be so rapid as to produce astrophysically relevant effects \cite{Duan:2010af, Wu:2017drk, Li:2021vqj, Ehring:2023abs, Ehring:2024mjx}. While there is no long-range interaction, neutrino collective behavior can be understood as dynamical flavor waves, so the theory of collective flavor conversions \hbox{\cite{Fiorillo:2024bzm, Fiorillo:2024uki, Fiorillo:2024pns, Fiorillo:2025ank}} is surprisingly similar to that of plasma waves.

In other words, an equivalent viewpoint to neutrino-neutrino refraction is that neutrinos source a coherent weak potential carrying flavor, acting as intermediary for refractive flavor exchange. These collective modes of the weak potential must therefore resemble waves in an electron-ion plasma. Under certain conditions, these flavor waves can turn unstable and grow spontaneously, triggering flavor conversion. We have previously identified this effect as resonant Cherenkov emission of such flavor waves \cite{Fiorillo:2024bzm}.

\begin{figure}
    \centering
    \includegraphics[width=0.37\columnwidth]{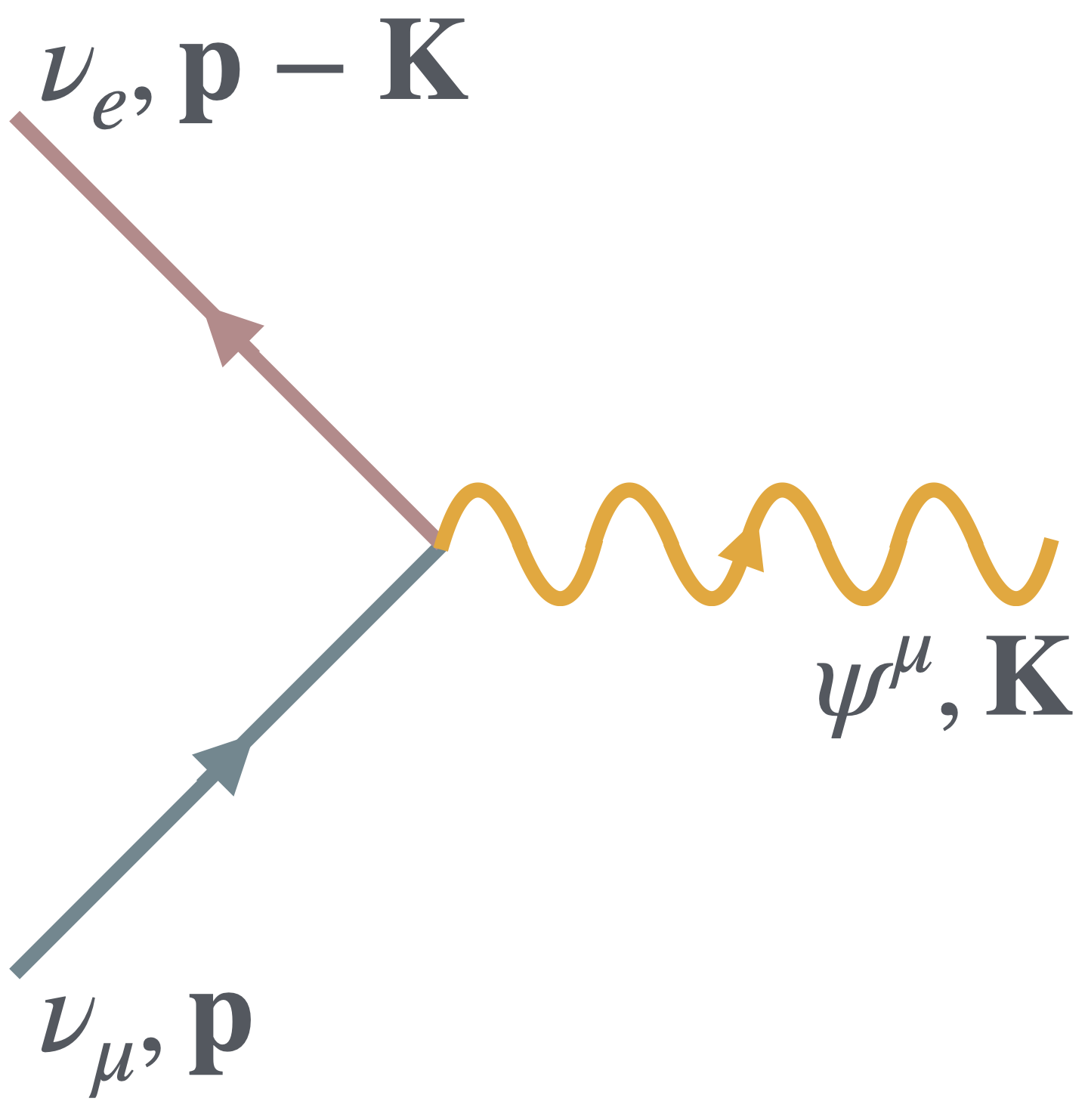}
    \vskip-6pt
    \caption{Elementary interaction vertex of the flavomon field. A $\nu_\mu$ (blue) converts into a $\nu_e$ (red), by emitting a flavomon (orange) which carries one unit of $\mu$ lepton number and minus one unit of $e$ lepton number.}
    \label{fig:fig1}
    \vskip-6pt
\end{figure}

The traditional description uses the neutrino degrees of freedom alone. The standard quantum kinetic equations \cite{Dolgov:1980cq, Rudsky, Sigl:1993ctk, Sirera:1998ia, Yamada:2000za, Vlasenko:2013fja, Volpe:2013uxl, Serreau:2014cfa, Kartavtsev:2015eva, Fiorillo:2024fnl, Fiorillo:2024wej} follow the flavor states of all individual neutrino modes. On the other hand, the analogy with collective plasma dynamics suggests that this description may be overly detailed. In a plasma, the collective motion of the entire system is described much more compactly by a new, emergent field, the plasmon. 

In this {\em Letter}, we introduce a similar description for the collective flavor dynamics of a collisionless neutrino gas. In place of the plasmon, which transports primarily energy and momentum, we introduce the flavor plasmon, or \textit{flavomon}, which also transports energy and momentum, but especially flavor. In a two-flavor picture, flavomons can be emitted by a $\nu_\mu$ turning into a $\nu_e$ through the elementary interaction vertex depicted in Fig.~\ref{fig:fig1}. In this sense, flavomons are similar to gluons, which can change the color of quarks. 

In this new paradigm, the small spatial and temporal scales associated with fast and slow flavor conversions are interpreted as wavelengths and periods of the flavomons produced in the instability. So, in principle, such small scales need not be resolved, in the same way that, e.g., the production of $\gamma$-rays does not require a spatial resolution comparable with their wavelength.

{\bf\textit{Interacting neutrinos and flavomons.}}---We study a dense gas of neutrinos of two flavors $i=e$ or $\mu$, where the weak interaction enables flavor exchange. For now we ignore antineutrinos and focus on the \textit{fast} limit, defined as the limit of vanishing neutrino masses; we will return to these points later. The interaction potential of this collisionless neutrino plasma is 
\begin{equation}\label{eq:Uint}
     \mathcal{U}=\frac{\sqrt{2}\GF}{2}\sum_{\overset{\bK,\bp,\bp'}{i,j}}
    a^\dagger_{i,\bp+\frac{\bK}{2}}a_{j,\bp-\frac{\bK}{2}} a^\dagger_{j,\bp'-\frac{\bK}{2}}a_{i,\bp'+\frac{\bK}{2}}
    \,v\cdot v'.
\end{equation}
Here $a_{i,\bp}$ is the annihilation operator for a neutrino with momentum $\bp$ and flavor $i$, $\GF$ the Fermi constant, and $v\cdot v'=v^\mu v'_\mu=1-\bv\cdot\bv'$ derives from the spinor contractions of the weak interaction in the near-homogeneous limit $|\bK|\ll|\bp|,|\bp'|$, which always applies in a realistic environment. (The complete interaction Hamiltonian includes additional flavor-blind terms that are explicitly shown in Ref.~\cite{Fiorillo:2024fnl}; they do not affect flavor dynamics and are therefore ignored.) We use the four-vector $v^\mu=(1,\bv)$, where $\bv=\bp/|\bp|$ is the neutrino velocity. Flavor exchange arises from off-diagonal terms under the sum of the type $a^\dagger_{e,\bp+\bK/2}a_{\mu,\bp-\bK/2}a^\dagger_{\mu,\bp'+\bK/2}a_{e,\bp'+\bK/2}$. Hence, an individual neutrino feels a refractive potential
\begin{equation}\label{eq:interaction_Hamiltonian}
    H_{\rm int}=\frac{\sqrt{2}\GF}{2} \sum_{\bp,\bK}\psi^\alpha_\bK v_\alpha a^\dagger_{\mu,\bp+\frac{\bK}{2}} a_{e,\bp-\frac{\bK}{2}}+\rm h.c.,
\end{equation}
using the field
\begin{equation}
    \psi^\alpha_\bK=2\sum_{\bp'}a^\dagger_{e,\bp'-\frac{\bK}{2}}a_{\mu,\bp'+\frac{\bK}{2}}v^{\prime\alpha}.
\end{equation}
The terms with $i=j$ in Eq.~\eqref{eq:Uint} introduce a $\nu_e$--$\nu_\mu$ energy splitting. For a test neutrino immersed in a homogeneous medium with $\nu_e$ and $\nu_\mu$ occupation numbers $n_{e,\bp}$ and $n_{\mu,\bp}$, the relevant Hamiltonian from Eq.~\eqref{eq:Uint} is
\begin{equation}\label{eq:H_split}
    H_{\rm split}=\sqrt{2}\GF\sum_\bp \left[n_e\cdot v\, a^\dagger_{e,\bp}a_{e,\bp}+n_\mu\cdot v\, a^\dagger_{\mu,\bp}a_{\mu,\bp}\right],
\end{equation}
where we have introduced the four-fluxes for the flavors $n_{e}^\alpha=\sum_\bp n_{e,\bp}v^\alpha$ and similarly for $n_\mu$.

In the quantum kinetic treatment, one studies the time evolution of the mean field $\langle\psi^\alpha\rangle$, a classical field that we now also call $\psi^\alpha$ and its Fourier transform $\psi^\alpha_{\bK,\Omega}$.
Its collective nature is evident because it involves a sum over all neutrinos in the medium. All its properties are encoded in a single function that we earlier introduced as flavor susceptibility~\cite{Fiorillo:2024bzm}, describing the reaction of the neutrino medium. If a flavor field is applied, the medium responds with a field of its own $\psi_{\rm med,\bK,\Omega}^\mu=\chi^\mu_\nu(\Omega,\bK) \psi^\nu_{\bK,\Omega}$, where $\Omega$ is the variation of the external field. Hence, in the absence of an external field, we must have
\begin{equation}\label{eq:dispersion_formal}
    \varepsilon^{\mu\nu}(\Omega,\bK)\psi_{\nu,\bK,\Omega}=0
\end{equation}
with $\varepsilon^{\mu\nu}=g^{\mu\nu}-\chi^{\mu\nu}$ and $g^{\mu\nu}$ the metric tensor. This consistency equation determines the allowed values of $\Omega$ for a given $\bK$, i.e., the flavor field can exhibit collective oscillations with the eigenfrequencies $\Omega(\bK)$ solving this dispersion relation. This picture is entirely classical, and underlies, implicitly or explicitly, all previous treatments of collective flavor evolution.

We now turn to a quantum treatment. If the collective field $\psi^{\alpha}$ can exhibit stationary oscillations, for now taken to be neither damped nor growing ($\mathrm{Im}\,\Omega=0$), there must be associated energy. Defining the field energy in a medium is notoriously tricky; while the total energy must be conserved, which part exactly to associate with the field alone is less obvious. 

However, we may follow the analogy with the electrodynamics of a dielectric medium, where a similar problem occurs. The dielectric function, here taken to be scalar, measures the displacement $\bD_{\bK,\Omega}=\varepsilon(\Omega,\bK)\bE_{\bK,\Omega}$. The energy of a monochromatic field (frequency $\Omega$) can be defined as~\cite{landau2013electrodynamics} (see also {\em Supplemental Material} \cite{supplementalmaterial})
\begin{equation}
    E_{\rm field}=\sum_\bK \frac{1}{2}\frac{\partial\left[\varepsilon(\Omega,\bK)\Omega\right]}{\partial\Omega} |\bE_{\bK,\Omega}|^2.
\end{equation}
Based purely on this analogy, we can therefore write the energy for the flavor field as
\begin{equation}
    E_{\rm field}= \frac{\sqrt{2}\GF}{4}\sum_\bK \psi^{\mu}_{\bK,\Omega}\psi^{\nu*}_{\bK,\Omega}\frac{\partial\left[\Omega \varepsilon_{\mu\nu}(\Omega,\bK)\right]}{\partial \Omega}.
\end{equation}
The overall factor is chosen such that in the absence of a medium response ($\varepsilon_{\mu\nu}=g_{\mu\nu}$), $E_{\rm field}$ equals the whole energy associated with $\psi^\mu_{\bK,\Omega}$ \cite{Fiorillo:2024bzm}. This expression is rigorously valid only for stationary oscillations. On the other hand, for modes that are weakly damped or weakly growing, such that $|\mathrm{Im}(\Omega)|\ll|\mathrm{Re}(\Omega)|$, which we shall generically call quasi-stationary modes, an independent $E_{\rm field}$ is still meaningful for practical purposes.

The field will therefore contain quasi-stationary components which we expand explicitly as
\begin{equation}
    \psi^\mu_\bK(t)= c_{\bK}e^\mu_{\bK} e^{-i\Omega_{\bK} t},
\end{equation}
where $e^\mu_{\bK}$ is the eigenvector and $\Omega_{\bK}$ the eigenfrequency associated with wavevector $\bK$, obtained from the dispersion relation Eq.~\eqref{eq:dispersion_formal}. It is understood that if, for a fixed value of $\bK$, the dispersion relation admits multiple eigenvalues, one must sum over all of them in the expansion. The energy can similarly be decomposed as
\begin{equation}\label{eq:classical_energy_field}
     E_{\rm field}=\frac{\sqrt{2}\GF}{4}\sum_{\bK} |c_{\bK}|^2\,\Omega_{\bK}
     \mathcal{Z}_{\bK}^{-1}
\end{equation}
where
\begin{equation}
      \mathcal{Z}_{\bK}^{-1}=e^\mu_{\bK}e^{*\nu}_{\bK}\partial_\Omega \varepsilon_{\mu\nu}(\Omega,\bK)\big|_{\Omega=\Omega_\bK}.
\end{equation}
We have used that the flavor dielectric function vanishes when applied to an eigenmode. 

The factor $\Omega_{\bK}\mathcal{Z}_{\bK}^{-1}$ accompanying $|c_{\bK}|^2$ need not be positive, i.e., there can be modes with negative energy, and indeed we find such cases in practical examples. This unintuitive behavior follows from dividing the energy between the collective modes and the neutrino medium and does not invalidate our approach. Negative energies are problematic only in equilibrium media. Our theory of flavomons is primarily of interest for non-equilibrium conditions, which can exhibit collective instabilities. In our {\em Supplemental Material} \cite{supplementalmaterial}, we show that for a medium in equilibrium, there is no fundamental inconsistency in the existence of negative-energy flavomons.

We now introduce the quanta of the flavor field by writing the energy of a given mode as a superposition of quanta. For positive-frequency components $\Omega_{\bK}>0$ and with $\mathcal{Z}_{\bK}>0$, the field energy is positive and expressed as $E_{\bK}=\Omega_{\bK} N_{\bK}$, where $N_{\bK}$ is the number of quanta. Comparison with Eq.~\eqref{eq:classical_energy_field} reveals that the coefficients $c_{\bK}$ can be reinterpreted as flavomon annihilation operators through the proportionality constant
\begin{equation}
    c_{\bK}=2 \sqrt{\frac{\mathcal{Z}_{\bK}}{\sqrt{2}\GF}}A_{\bK}.
\end{equation}
Here $A^\dagger_{\bK}$ and $A_{\bK}$ are the creation and annihilation operators for flavomons, and \smash{$N_{\bK}=A^\dagger_{\bK}A_{\bK}$} their number operator. We recognize $\mathcal{Z}_{\bK}$ as the wavefunction renormalization of the field induced by the medium. This is well known from quantum field theory, as $\varepsilon_{\mu\nu}$ plays a role analogous to the flavomon self-energy. 

If $\Omega_{\bK}>0$ and $\mathcal{Z}_{\bK}<0$, the field energy is negative and so the quanta have negative energy. The coefficients $c_{\bK}$ are now anti-flavomon creation operators
\begin{equation}
    c_{\bK}=2 \sqrt{\frac{-\mathcal{Z}_{\bK}}{\sqrt{2}\GF}}B^\dagger_{\bK},
\end{equation}
so that $E_{\bK}=-\Omega_{\bK}B^\dagger_{\bK}B_{\bK}$. For $\Omega_{\bK}<0$, we have instead that $\mathcal{Z}_{\bK}>0$ is associated with flavomons created by $c_{\bK}$, while $\mathcal{Z}_{\bK}<0$ with anti-flavomons annihilated by $c_{\bK}$.
Notice that flavomons and antiflavomons do not generally appear symmetrically, since a solution with $\Omega_{\bK}$ does not imply the conjugate one with $-\Omega_{\bK}$.

We can finally replace in Eq.~\eqref{eq:interaction_Hamiltonian} to obtain the neutrino-flavomon interaction,
\begin{equation}\label{eq:interaction_Hamiltonian-QFT}
    H_{\rm int}=\sum_{\bp,\bK}\sqrt{\sqrt{2}\GF |\mathcal{Z}_{\bK}|}(e_{\bK}\cdot v) A_{\bK}a^\dagger_{\mu,\bp+\frac{\bK}{2}}a_{e,\bp-\frac{\bK}{2}}.
\end{equation}
For simplicity we have left out the terms corresponding to antiflavomons. The neutrino-flavomon coupling is proportional to $\sqrt{\GF}$, directly signaling that the neutrino-flavomon dynamics happens over timescales determined by $\GF$, not $\GF^2$.

{\bf\textit{Kinetic equations for neutrinos and flavomons.}} To lowest order in $\GF$, flavomons are produced by the decay $\nu_\mu\to\nu_e\psi$ and absorbed by its inverse. We first consider $\nu_\mu(\bp)\to\nu_e(\bp')+\psi_{\bK}$, and find
\begin{eqnarray}
    \frac{\partial N_{\bK}}{\partial t}
        &=&\int \frac{d^3\bp}{(2\pi)^3}\int \frac{d^3\bp'}{(2\pi)^3}
    (2\pi)^4\delta^{(4)}(p_\mu-p'_\mu-K_\mu)
    \nonumber\\[1ex]
    &&\kern1em{}\times|\mathcal{M}|^2 n_{\mu,\bp}(1-n_{e,\bp'})(1+N_{\bK}),
\end{eqnarray}
where $p_\mu$, $p'_\mu$, and $K_\mu$ are the four-momenta of $\nu_\mu$, $\nu_e$, and $\psi$ respectively. The spatial delta functions can be used to integrate out $\bp'$, whereas the remaining one imposes energy conservation. The energy of the initial $\nu_\mu$ is $p^0=|\bp|+\sqrt{2}\GF n_\mu\cdot v$, where the last term is the refractive energy shift in Eq.~\eqref{eq:H_split}, while the energy of the final $\nu_e$ is $p^{\prime\,0}=|\bp-\bK|+\sqrt{2}\GF n_e\cdot v$. Therefore, the condition of energy conservation, expanded for $|\bK|\ll |\bp|$, is
\begin{equation}
    \Omega_\bK-\bv\cdot \bK+\sqrt{2}\GF G\cdot v=0,
\end{equation}
where $G^\alpha=n^\alpha_e-n^\alpha_\mu$. This can be written in covariant form with the four-vector $k^\alpha=K^\alpha+\sqrt{2}G_F G^\alpha$ as $k\cdot v=0$. Finally, using the interaction Hamiltonian in Eq.~\eqref{eq:interaction_Hamiltonian-QFT},
\begin{equation}
    |\mathcal{M}|^2=\sqrt{2}\GF |\mathcal{Z}_{\bK}||e_{\bK}\cdot v|^2
\end{equation}
is the squared matrix element. We assume here $\mathcal{Z}_{\bK}>0$; if $\mathcal{Z}_{\bK}<0$, the state corresponds to an anti-flavomon and therefore the relevant decay is $\nu_e\to\nu_\mu\overline{\psi}$.

The total rate of change in the flavomon number must include also the rate for flavomon absorption $\nu_e\psi\to\nu_\mu$, so we finally obtain the rate equation
\begin{eqnarray}\label{eq:kinetic_flavomon}
    \kern-1em&&\frac{\partial N_\bK}{\partial t}=\sqrt{2}\GF\int \frac{d^3\bp}{(2\pi)^3}2\pi\delta(k\cdot v)|\mathcal{Z}_{\bK}| |e_{\bK}\cdot v|^2
    \\ \nonumber 
    \kern-1em&&{}\times 
    \bigl[n_{\mu,\bp}(1-n_{e,\bp-\bK})(1+N_\bK)-n_{e,\bp-\bK}(1-n_{\mu,\bp})N_\bK\bigr].
\end{eqnarray}
The term in square parenthesis can now be approximated in the classical limit $N_{\bK}\gg 1$ and $|\bK|\ll |\bp|$ to give $\partial N_{\bK}/\partial t=2\gamma_{\bK}N_{\bK}$ with
\begin{equation}
    \gamma_{\bK}=-\pi\int \frac{d^3\bp}{(2\pi)^3}\delta(k\cdot v)\frac{G_\bp |e_{\bK}\cdot v|^2}{\partial_\Omega\varepsilon_{\mu\nu}e_{\bK}^\mu e_{\bK}^{*\nu}},
\end{equation}
where $G_\bp=n_{e,\bp}-n_{\mu,\bp}$. 

To understand better this expression, we notice that the flavor dielectric function can be explicitly determined from the equations of motion to be~\cite{Fiorillo:2024bzm}
\begin{equation}
    \varepsilon^{\mu\nu}=g^{\mu\nu}-\int \frac{d^3\bp}{(2\pi)^3}\frac{G_\bp v^\mu v^\nu}{k\cdot v+i\epsilon},
\end{equation}
which allows us to rewrite
\begin{equation}
    \gamma_{\bK}=-\frac{\mathrm{Im}(\varepsilon_{\mu\nu})e^\mu_{\bK}e^{*\nu}_{\bK}}{\partial_\Omega \varepsilon_{\mu\nu}e_{\bK}^\mu e_{\bK}^{*\nu}}.
\end{equation}
This result coincides with the growth rate of weakly unstable (or weakly damped) modes derived in Ref.~\cite{Fiorillo:2024uki}, confirming that fast flavor instabilities correspond to stimulated decay $\nu_\mu\to\nu_e\psi$. For neutrinos, we find analogously
\begin{eqnarray}
    \kern-1em&&
    \frac{\partial n_{e,\bp}}{\partial t}=-\frac{\partial n_{\mu,\bp}}{\partial t}\\ \nonumber 
    \kern-1em&&{}=
    \int \frac{d^3\bK}{(2\pi)^3}2\pi \delta(k\cdot v)\sqrt{2}\GF 
    |\mathcal{Z}_{\bK}|\,|e_{\bK}\cdot v|^2\\ \nonumber 
    \kern-1em&&
    {}\times\bigl[n_{\mu,\bp}(1-n_{e,\bp-\bK})(1+N_\bK)-n_{e,\bp-\bK}N_\bK (1-n_{\mu,\bp})\bigr].
\end{eqnarray}
These equations are \textit{not} equivalent to the linear theory of fast instabilities, in which the $\nu_e$ and $\nu_\mu$ number densities are taken to be fixed. Instead, they are equivalent to the quasi-linear theory of fast instabilities that we derived in Ref.~\cite{Fiorillo:2024fnl} for the case of discrete neutrino beams. Recall that $N_\bK\propto|\psi|^2$, where $\psi$ corresponds to the off-diagonal element of the flavor density matrix. In other words, the quasi-linear equation is linear in the large diagonal elements, but quadratic in the small off-diagonals.

In any practical situation, flavomons are highly occupied ($N_\bK\gg1$) and the kinetic equation is classical. While the quantum treatment may thus seem like a distraction, it actually provides the most transparent and direct path to deriving a kinetic equation for classical waves. In addition, it provides a true quantum feature, the spontaneous flavomon production rate. While quantum fluctuations to seed flavor instabilities were previously discussed \cite{Sawyer:2022ugt}, we here provide an explicit rate, that is however very small. The dominant trigger probably remains forcing by vacuum masses and flavor mixing, which are not part of the fast flavor paradigm.

The new kinetic equations automatically conserve both the number of neutrinos and the total lepton number. The latter, however, is only conserved when the flavomon contribution is included
\begin{equation}
    \frac{\partial}{\partial t}\left[\int \frac{d^3\bp}{(2\pi)^3}n_{e,\bp}-\int \frac{d^3\bK}{(2\pi)^3}N_{\bK}\right]=0,
\end{equation}
and similarly for the muon lepton number.

The kinetic equations are written here for a homogeneous setup. However, it is straightforward to generalize them to include neutrino and flavomon streaming terms, and therefore can describe systems with large-scale inhomogeneities. The relevance of streaming terms has been empirically emphasized by the NBI group
\cite{Shalgar:2022lvv, Shalgar:2022rjj, Cornelius:2023eop, Cornelius:2024zsb}, and we have recently proven their necessity~\cite{Fiorillo:2025ank}.  

We also stress that these kinetic equations naturally introduce some form of coarse graining. They are based on a particle-physics computation of the decay rate that is valid only if the distribution is roughly homogeneous on the Compton length scales. This is microscopic for neutrinos, but for flavomons, it is the wavelength of the unstable modes of the order \smash{$(\mu\epsilon)^{-1}=\bigl[\sqrt{2}\GF (n_\nu-n_{\overline{\nu}})\bigr]^{-1}$}, which is macroscopic. Therefore, the neutrino distributions are smeared over large distances. Smaller scale fluctuations in the neutrino occupation numbers enter only at higher orders of perturbation theory, as we discuss in more detail in the {\em Supplemental Material}~\cite{supplementalmaterial}.

{\bf\textit{Discussion.}}---We have derived new kinetic equations for neutrinos and flavomons, the quanta of the traditional flavor waves. The formal starting point was our previously identified flavor susceptibility $\chi^{\mu\nu}$ that encodes all information of the interacting neutrino gas on the level of linear response theory. It provides the flavomon propagator, the dispersion relation through its poles, and the wavefunction renormalization factor from their residues, and thus the conversion factor from a classical flavor-wave amplitude to the corresponding number of quanta. In the new paradigm, the classical flavor instability of the usual quantum kinetic equations corresponds to the occupation number of flavomons growing exponentially because of stimulated decay $\nu_\mu\to\nu_e\psi$.

The dynamics of this process is similar to a laser, an analogy already used to describe plasma instabilities. In our case of $\nu_\mu\to\nu_e\psi$, in a $\nu_e$ dominated medium, the dominance of $\nu_\mu$ along certain directions (angular crossings) leads to their stimulated decay until equipartition along that direction is reached and the right-hand side of Eq.~\eqref{eq:kinetic_flavomon} vanishes. Thus, a simple application of detailed balance explains the role of the angular crossing \cite{Morinaga:2021vmc, Dasgupta:2021gfs, Fiorillo:2024bzm}, as well as its removal that was empirically observed in homogeneous systems \cite{Nagakura:2022kic, Zaizen:2022cik}, and theoretically proven for a simplified case~\cite{Fiorillo:2024qbl}. 

In addition, flavomon emission changes neutrino energies, explaining with simple kinematics the previously shown nonconservation of refractive energy \cite{Fiorillo:2024fnl}. The flavomon picture naturally explains what are arguably the most fundamental features of fast flavor conversions that have been understood in the last decade. This framework also naturally describes ``random phase'' ensembles in which the neutrino
behavior is not regular. A direct counterexample is the fast pendulum \cite{Padilla-Gay:2021haz, Johns:2019izj, Fiorillo:2023mze}, which itself is a sub-class of flavor solitons \cite{Fiorillo:2023hlk}. Such solutions are macroscopic, classical configurations of the flavomon field, rather than an ensemble of flavomons, in the same way that solitons in a quantum field theory correspond to classical field configurations rather than ensembles of particles.

At a formal level, our kinetic equations are limited by mainly two approximations. First, we have accounted only for on-shell flavomons, showing up in the energy-conserving delta function. This should more correctly be replaced by a spectral function which accounts for the broadening of the resonance due to the flavomon growth (or damping) rate $\gamma_{\bK}$. Thus, if an angular crossing is present, with a small ``flipped'' region of
$\nu_\mu$ dominance which is led to equipartition by flavomon emission, even neutrinos in the $\nu_e$ dominated region can interact with these flavomons nonresonantly. One consequence is that the muon lepton number that is removed from the flipped region, which in our equations is entirely transferred to flavomons, can migrate via nonresonant absorption to the unflipped region.

Our second approximation is to ignore nonlinear interaction among flavomons. This is justified while their number is not too large and certainly true in the early stages of the instability, but remains an open question at later stages. Our recent finding that for weak instabilities the produced flavomons always leave their production region~\cite{Fiorillo:2025ank}, as well as the general insight that, once an instability appears, the medium reacts in the opposite direction to remove that condition, both suggest that the number of flavomons might not grow too large. However, such an assumption should be tested. We emphasize that the neglect of both nonresonant interactions and nonlinear flavomon interactions can be investigated systematically through the methods introduced here. 

We have focused on fast flavor conversions, but slow ones can be included. Mass splittings alter the kinematic condition for $\nu_\mu\to\nu_e\psi$ by an additional energy splitting that depends on the neutrino energy. An angular crossing is then not necessary, an energy crossing is enough. We have also left out matter refraction, that provides an additional shift of the flavomon energy, possibly affecting their large-scale propagation. We have also ignored collisions, which might damp flavomons or potentially produce them via collisional instabilities \cite{Johns:2021qby, Xiong:2022zqz, Liu:2023pjw, Lin:2022dek, Johns:2022yqy, Padilla-Gay:2022wck, Fiorillo:2023ajs}; however collisions do not introduce any fundamental new difficulty, as they simply modify the dispersion relation of the flavomons. 

We have considered only neutrinos, but including antineutrinos is straightforward, since they simply carry opposite lepton number. Thus, in a $\overline\nu_e$ system, decays of the form $\overline{\nu}_e\to\overline{\nu}_\mu \psi$ will occur. Finally, we have restricted our treatment to the two-flavor case, in which there is only one (complex) flavomon field, carrying a lepton charge $e\overline{\mu}$. This corresponds to the two off-diagonal mediator of the SU(2) symmetry of the theory. In the three-flavor case, the symmetry is expanded to SU(3), so there will be three complex flavomon fields, which carry charges $e\overline{\mu}$, $e\overline{\tau}$, and $\mu\overline{\tau}$. However, no other fundamental changes to the theory are introduced by the three-flavor case.

The main advantage of the concept of flavomons is that it naturally circumvents the main challenge of collective flavor conversions, i.e., the short distances and small timescales involved. These are now the wavelengths and periods of flavomons, and thus need not be resolved. The only scale of physical interest is the growth rate of the instability, and realistically this will never become too large, since the instability is slowly driven by the medium reacting over collisional timescales. Whether the flavomon picture will ultimately allow one to circumvent entirely these problem hinges on the critical question of nonlinear interactions. If their number grows so large that their nonlinear interaction is more rapid than their periods, then a new framework is needed. Luckily, this question can be investigated within the flavomon framework itself, by testing in realistic circumstances, i.e., accounting for the medium evolution and large-scale variations allowing for flavomon and neutrino propagation, whether this picture remains valid.

{\bf\textit{Acknowledgments.}}---DFGF is supported by the Ale\-xander von Humboldt Foundation (Germany), whereas GGR acknowledges partial support by the German Research Foundation (DFG) through the Collaborative Research Centre ``Neutrinos and Dark Matter in Astro- and Particle Physics (NDM),'' Grant SFB-1258\,--\,283604770, and under Germany’s Excellence Strategy through the Cluster of Excellence ORIGINS EXC-2094-390783311.

\bibliographystyle{bibi}
\bibliography{References}
\onecolumngrid

\include{SMmod.tex}

\end{document}

%% file: SMMod.tex
\onecolumngrid
\appendix

\setcounter{equation}{0}
\setcounter{figure}{0}
\setcounter{table}{0}
\setcounter{page}{1}
\makeatletter
\renewcommand{\theequation}{S\arabic{equation}}
\renewcommand{\thefigure}{S\arabic{figure}}
\renewcommand{\thepage}{S\arabic{page}}

\begin{center}
\textbf{\large Supplemental Material for the Letter\\[0.5ex]
{\em Collective flavor conversions are interactions of neutrinos\\ with quantized flavor waves}}
\end{center}

\bigskip

In this Supplemental Material, we recapitulate how plasmons are conventionally obtained as quanta of density fluctuations in a collisionless plasma, and show the formal mapping for the case of the flavomon. We also introduce flavomons directly from the quantum field theory of interacting neutrinos, and determine the flavomon properties for an example distribution which has a fast instability. Finally, we discuss the properties of flavomons as excitations in a medium in thermal and chemical equilibrium.

\bigskip

\twocolumngrid

\section{A.~Plasmons as quantized electrostatic waves.}

The focus of this work is to find a definition of flavor plasmons (flavomons) as quanta of collective flavor waves supported by a neutrino medium. Such a description is completely analogous to the definition of plasmons as the quanta of the electric field in collective plasma waves, an analogy that has inspired our recent theoretical works on this topic \cite{Fiorillo:2024bzm, Fiorillo:2024uki, Fiorillo:2024pns, Fiorillo:2025ank}. We here review how the definition of plasmons emerges from the classical theory of a collisionless plasma, focusing on longitudinal plasmons, and is meant to familiarize the flavor community with the parallel concepts in plasma physics. In this sense, our present discussion is complementary to the Appendix of Ref.~\cite{Fiorillo:2024bzm}, in which the plasmon was obtained as a renormalized excitation of the electrostatic potential using quantum field theory. Here we start instead from the classical theory of plasma to reach the quantum definition. This exposition will allow for a direct mapping to flavor waves, which are also originally defined as a classical field.

The plasma properties are completely defined by the distribution function $f_\bp(\br,t)$ of electrons with momentum $\bp$ at point $\br$ and time $t$. They produce an electric field which satisfies the first Maxwell equation
\begin{equation}
    \boldsymbol{\nabla}\cdot \bE=\rho,
\end{equation}
where $\rho=\sum_\bp e f_\bp$ is the charge density, with $e$ the unit electron charge. (We use rationalized Lorentz--Heaviside units, where no $4\pi$ appears on the right-hand side of Gauss' Law, and where $\alpha=e^2/4\pi$, as well as natural units with $\hbar=c=1$.) For a medium the total charge can be written in the usual form $\rho=-\boldsymbol{\nabla}\cdot \bP$, where $\bP$ is the vector of  polarization, so that the equation can be more compactly written as
\begin{equation}
    \boldsymbol{\nabla}\cdot \bD=0,
\end{equation}
where $\bD=\bE+\bP$ is the displacement vector.

The electric field $\bE$, and in particular its longitudinal component, ultimately describes longitudinal plasma waves associated with density oscillations. The question is how to define a quantum of such excitations, which can be seen as a normalization problem in the sense of how many quanta (in a coherent state) are needed to achieve a specified classical field strength. The strategy that is conventionally adopted, following the early days of quantum mechanics, is to define an energy for the field, and express it as the sum of the contributions of quantized states with a given frequency. 

Therefore, the very first step is to define the energy of plasma waves. This is not a trivial task, and is not always possible. In a dispersive medium such as a plasma, the electric field continuously exchanges energy with the electrons. Because of this dynamics, plasma waves can be Landau-damped, dissipating energy into the medium electrons, or unstable, extracting energy from them. However, for those waves which are only weakly Landau-damped or unstable, the wave energy can still be approximately defined, since such energy gains or losses are very slow.

The usual starting point to define the medium energy is the rate at which the electric field dissipates its energy onto an external charge configuration. If an external current density $\delta \bj$ passes through the medium, the energy dissipated per unit time is
\begin{equation}\label{eq:dW-plasma}
    \delta W=\delta\bj\cdot\bE.
\end{equation}
This is a well-known formula, yet it is illuminating to derive it from first principles, because the analog for a flavor wave, namely the energy dissipated in the presence of an external neutrino, has never been examined. To derive Eq.~\eqref{eq:dW-plasma}, we note that the external electrons (momentum $\bp$) forming the current $\delta \bj$ obey Newton's equation
\begin{equation}
    \frac{d\bp}{dt}=e\bE.
\end{equation}
Therefore, the energy gained by each of them per unit time is $e\bE\cdot \bv$. The total energy gained by the external electrons, which have a distribution function $\delta f_\bp$, is
\begin{equation}
    \delta W=\sum_\bp \delta f_\bp e \bE \cdot \bv=\delta\bj\cdot \bE,
\end{equation}
where $\delta\bj=\sum_\bp \delta f_\bp e \bv$.

To proceed, we note that $\delta \bj$ must satisfy the fourth Maxwell equation
\begin{equation}
    \boldsymbol{\nabla}\times \bB=\delta \bj+\frac{\partial \bD}{\partial t}.
\end{equation}
We neglect the magnetic field, which is responsible for time-dependent effects associated with spatial transport of electromagnetic energy, and conclude that
\begin{equation}
    \delta W=-\bE\cdot\frac{\partial\bD}{\partial t}.
\end{equation}
This is the energy absorbed by the medium, so the energy gained by the field per unit time is
\begin{equation}
    \partial_t E_{\rm field}=-\int d^3\br\, W=\int d^3\br\,\bE\cdot\frac{\partial\bD}{\partial t}.
\end{equation}
The information about the medium properties is now completely encoded in the relation between $\bD$ and $\bE$. If the medium can be considered linear, meaning that the displacement is a linear response to the electric field, depending on frequency and wavenumber, all the information is in the dielectric tensor $\varepsilon_{ij}(\Omega,\bK)$ in Fourier space. 
Denoting quantities in Fourier space with the same letters as in coordinate space we thus have
\begin{equation}\label{eq:response}
{D}_i(\Omega,\bK)=\varepsilon_{ij}(\Omega,\bK) {E}_j(\Omega,\bK).
\end{equation}
In coordinate space, this linear relation becomes a convolution, representing the phenomena of spatial and temporal dispersion.
This convolution is often symbolically written in the form 
$\bD=\hat{\varepsilon}\,\bE$ (see, e.g., Ref.~\cite{landau2013electrodynamics}) so that
\begin{equation}
    \partial_t E_{\rm field}=\int d^3\br\, \bE\cdot\frac{\partial(\hat{\varepsilon}\bE)}{\partial t}.
\end{equation}
By integrating over time from $-\infty$ to $+\infty$, we obtain the total energy $Q=\int_{-\infty}^{+\infty} dt\,\partial_t E_{\rm field}$ gained by the field
\begin{equation}
    Q=\sum_{\bK}\int \frac{d\Omega}{2\pi}{E}_i^*(\Omega,\bK){E}_j(\Omega,\bK)\left[-i\Omega \varepsilon_{ij}(\Omega,\bK)\right].
\end{equation}
This can also be written as
\begin{equation}
    Q=\frac{1}{2}\sum_\bK \int \frac{d\Omega}{2\pi}{E}_i^*{E}_ji\Omega\left[-\varepsilon_{ij}+\varepsilon^*_{ji}\right],
\end{equation}
recovering the well-known result that energy dissipation is associated with the imaginary part of the dielectric function, which in a plasma describes Landau damping or instability.

Integrating from $-\infty$ to $t$ instead provides the energy that the field has extracted up to that time, which can be interpreted as the field energy itself, and we find
\begin{eqnarray}
    \kern-3em
    E_{\rm field}&=&\frac{1}{2}\sum_\bK \int \frac{d\Omega}{2\pi}\frac{d\Omega'}{2\pi}{E}_i(\Omega',\bK){E}_j(\Omega,\bK)
    \nonumber\\[1.5ex]
    \kern-3em&&{}\times\frac{\Omega \varepsilon_{ij}(\Omega,\bK)-\Omega' \varepsilon^*_{ji}(\Omega',\bK)}{\Omega-\Omega'}\,e^{-i(\Omega-\Omega')t}.
\end{eqnarray}
If the field is on average time-independent, the ensemble average $\langle {E}_i(\Omega',\bK){E}_j(\Omega,\bK)\rangle=2\pi\delta(\Omega-\Omega')P_{ij}(\Omega,\bK)$. If dissipation is weak, as we assumed from the beginning, we can take $\varepsilon_{ij}(\Omega,\bK)=\varepsilon^*_{ji}(\Omega,\bK)$, and finally have
\begin{equation}
    E_{\rm field}=\frac{1}{2}\sum_\bK \int \frac{d\Omega}{2\pi}P_{ij}(\Omega,\bK)\partial_\Omega\bigl[\Omega \varepsilon_{ij}(\Omega,\bK)\bigr].
\end{equation}
This is the energy of the electric field. 

Writing it as a linear combination of eigenmodes with frequencies $\Omega_\bK$ implies
${E}_i(\Omega,\bK)={E}_{i,\bK}2\pi \delta(\Omega-\Omega_\bK)$ so that
$P_{ij}={E}_{i,\bK}^*{E}_{j,\bK}2\pi \delta(\Omega-\Omega_\bK)$ and finally
\begin{equation}
    E_{\rm field}=\frac{1}{2}\sum_\bK {E}^*_{i,\bK}{E}_{j,\bK}\Omega_\bK \left[\partial_\Omega \varepsilon_{ij}(\Omega,\bK)\right]_{\Omega=\Omega_\bK},
\end{equation}
where we have used that by definition for eigenoscillations $\varepsilon_{ij}(\Omega_\bK,\bK)E_{j,\bK}=0$.

With this result, the electric field can be expressed as a superposition of the creation and annihilation operators of quanta
\begin{eqnarray}
    \bE&=&\sum_\bK \sqrt{\frac{2}{\left[\partial_\Omega \varepsilon_{ij}(\Omega,\bK)\right]_{\Omega=\Omega_\bK} e_{i,\bK}e_{j,\bK}^*}}
    \\ \nonumber
    &&\kern7em{}\times\be_\bK a_\bK e^{-i\Omega_\bK t+i\bK\cdot\br}
    +\hbox{h.c.}
\end{eqnarray}
Here $a_\bK$ is the annihilation operator of a plasmon. Its polarization vector must satisfy the eigenvector equation $\varepsilon_{ij}(\Omega_\bK,\bK)e_{j,\bK}=0$. In the denominator, the factor $\left[\partial_\Omega \varepsilon_{ij}(\Omega,\bK)\right]_{\Omega=\Omega_\bK} e_{i,\bK}e_{j,\bK}^*$ can be interpreted as the 
wavefunction renormalization, familiar from quantum field theory. Indeed, the dielectric function $\varepsilon_{ij}$ is connected with the inverse propagator of a photon, and therefore its derivative with respect to $\Omega$, evaluated at the eigenfrequency, corresponds to the residue of the pole.

With this definition, we see that the field energy is the usual sum over the energies of quantum oscillators, i.e.,
\begin{equation}
    E_{\rm field}=\sum_\bK \Omega_\bK a^\dagger_\bK a_\bK.
\end{equation}
This well-known result confirms that the definition of the plasmon as a quantum excitation of the medium is consistent. Mapping this line of reasoning to flavor waves will next allow us to identify the flavor plasmons (flavomons) as quantized excitations with the correct amplitude normalization.

\section{B.~Energy of the flavor waves.}

The definition of the energy associated with flavor waves is subtle. They represent collective motions of the medium, similar to plasma waves, so it is once more unclear which fraction of the total energy to associate with the collective motion itself and which fraction instead with individual particles. Such a separation is not rigorously possible because Landau damping and instabilities spawn the continuous exchange of energy between the flavor field and individual neutrinos. However, for a weakly unstable neutrino gas, one can approximately identify a flavor wave as an independent entity, which is at the core of this work, and analogous to similar questions for plasma waves. This separation can only be done in the so-called transparency ranges, namely the ranges of frequency where energy dissipation of collective waves is sufficiently weak. 

To define the energy in these frequency ranges, we follow the same procedure that is usually adopted for the definition of the electromagnetic field energy in a plasma. The main physical idea is to introduce a test particle, and to determine the amount of energy that it exchanges with the field per unit time. This provides an operational definition of the field energy itself. Therefore, we assume that the neutrino medium is described by a density matrix $\rho_\bp=D_\bp+\Delta_\bp$. Here
\begin{equation}
    D_\bp=\frac{n_{\bp}}{2}+\frac{G_\bp}{2}\sigma_z
\end{equation}
is the unperturbed density matrix, which by definition does not have small-scale inhomogeneities and only counts the space-averaged number of electron and muon neutrinos. Moreover, $G_\bp$ is the spectrum, i.e., the initial $z$-component of the polarization vector, which itself is $\vec{D}_\bp=G_\bp \vec{e}_z$. On the other hand, $\Delta_\bp$ is the fluctuating part of the density matrix associated with the medium, i.e., the ensemble of the flavor waves, which we write as $\Delta_\bp=\vec{\Delta}_\bp\cdot \vec{\sigma}/2$. For fast flavor waves, energy does not appear in the EOM, so that often one uses energy-integrated quantities. Instead, here we consider flavor waves depending on momentum $\bp$.

We now introduce a test neutrino, whose density matrix is written as
\begin{equation}
    \delta \rho_\bp=\frac{\delta n_\bp}{2}+\frac{\delta \vec{P}_\bp \cdot \vec{\sigma}}{2}.
\end{equation}
We next introduce a generalized notation that makes the rest of the calculations more convenient, by defining four-vector and four-tensor fields associated with a density matrix as
\begin{equation}
    \rho^\mu=\sum_\bp \rho_\bp v^\mu
    \quad\hbox{and}\quad
    \rho^{\mu\nu}=\sum_\bp \rho_\bp v^\mu v^\nu,
\end{equation}
where $v^\mu=(1,\bv)$ is the four-velocity for a neutrino with momentum $\bp$. By definition, we have $\bv=\partial \varepsilon_\bp/\partial \bp$, where $\varepsilon_\bp$ is the neutrino kinetic energy 

The Hamiltonian of the test neutrino contains both the kinetic energy $\varepsilon_\bp$ and the refractive potential energy, so that $\Omega_\bp=\varepsilon_\bp+\sqrt{2}\GF v^\mu \left[\rho_\mu+\mathrm{Tr}(\rho_\mu)\right]$~\cite{Fiorillo:2024fnl}, whereas there is of course no refractive interaction of the test neutrino with itself. We can neglect in this expression any term depending on the trace of $\rho_\bp$, which do not affect the flavor dynamics, so that
\begin{equation}
    \Omega_\bp=\varepsilon_\bp+\frac{\sqrt{2}\GF v^\mu}{2}\bigl(\vec{D}_\mu+\vec{\Delta}_\mu\bigr)\cdot \vec{\sigma}.
\end{equation}
The individual neutrino energy, i.e., the part not associated with the interaction with the fluctuating flavor field, is therefore
\begin{equation}
    \Omega^0_\bp=\varepsilon_\bp+\frac{\sqrt{2}\GF v^\mu}{2}\vec{D}_\mu\cdot \vec{\sigma}.
\end{equation}
The general kinetic equation for the test neutrino becomes
now~\cite{Fiorillo:2024fnl}
\begin{widetext}
\begin{equation}
    \partial_t \delta\rho_\bp+\frac{1}{2}\left\{\partial_{p_i} \Omega_\bp,\partial_i\delta\rho_\bp\right\}-\frac{1}{2}\left\{\partial_{i} \Omega_\bp,\partial_{p_i}\delta\rho_\bp\right\}=-i[\Omega_\bp,\delta\rho_\bp].
\end{equation}
These equations can be split into a trace part, describing how the neutrino density changes in response to the flavor field, and a traceless part, describing the flavor evolution. For the trace part, we find
\begin{equation}
    \partial_t \delta n_\bp+\partial_i\left[\delta n_\bp v^i+\frac{\sqrt{2}\GF}{2}\delta\vec{P}_\bp \cdot  \bigl(\vec{D}_\mu+\vec{\Delta}_\mu\bigr) \partial_{p_i}v^\mu\right]-\frac{\sqrt{2}\GF}{2}\partial_{p_i}\left[\delta \vec{P}_\bp\cdot \partial_i\vec{\Delta}_\mu v^\mu\right]=0.
\end{equation}
The term in the first square bracket describes a spatial particle
flow, while the second square bracket a particle flow in momentum space, and therefore corresponds to a net change of neutrino kinetic energy~\cite{Fiorillo:2024fnl}.

The traceless part of the equations of motion describes the flavor evolution. We find
\begin{equation}\label{eq:test_neutrino_polarization}
    \partial_t \delta \vec{P}_\bp+\partial_i(\delta \vec{P}_\bp v^i)=\sqrt{2} \GF v^\mu(\vec{D}_\mu+\vec{\Delta}_\mu)\times \delta \vec{P}_\bp,
\end{equation}
where only terms of first order in $\sqrt{2}\GF$ have been kept.

We now proceed to obtain the amount of energy that the test neutrino can take from, or give to, the flavor field $\Delta^\mu$. We first identify the part of the test neutrino energy not associated with the interaction with the flavor field; this must be defined as $\delta U=\sum_\bp \mathrm{Tr}(\Omega^0_\bp \delta \rho_\bp)$, so
\begin{equation}
    \delta U=\sum_\bp\left[\varepsilon_\bp \delta n_\bp+\frac{\sqrt{2}\GF v^\mu \vec{D}_\mu \cdot \delta \vec{P}_\bp}{2}\right]
    =\sum_\bp \varepsilon_\bp \delta n_\bp+\frac{\sqrt{2}\GF\vec{D}_\mu\cdot \delta \vec{P}^\mu}{2}. 
\end{equation}
We now differentiate this expression with respect to time and use the EOMs to reduce it to the form
\begin{equation}\label{eq:energy_balance}
    \partial_t \delta U+\partial_i \delta\Phi^i=\frac{\sqrt{2}\GF}{2}\vec{\Delta}_\mu \cdot \partial_i\delta\vec{P}^{i\mu}+\GF^2\vec{D}_\mu \cdot(\vec{\Delta}_\nu\times \delta\vec{P}^{\mu\nu})=\delta W,
\end{equation}
where
\begin{equation}
    \delta \Phi^i=\sum_\bp \delta n_\bp \varepsilon_\bp v^i+\frac{\sqrt{2}\GF}{2}\sum_\bp \delta\vec{P}_\bp \cdot (\vec{D}_\bp+\vec{\Delta}_\bp)\partial_{p_i}v^\mu \varepsilon_\bp+\frac{\sqrt{2}\GF}{2}(\vec{D}_\mu+\vec{\Delta}_\mu)\cdot\delta\vec{P}^{i\mu}.
\end{equation}
\end{widetext}
Equation~\eqref{eq:energy_balance} is now in the form of an energy balance equation; the term $\partial_i \delta\Phi^i$ describes the spatial transport of energy, while the right-hand side $\delta W$ is the energy exchanged per unit time between the test neutrino and the field in the medium, playing the same role as the power transferred to a test charge by an electric field $\delta W=\delta \bf j\cdot E$, where $\delta \bf j$ is the test charge current and $\bf E$ the electric field. To move further, we notice that this can be rewritten as
\begin{equation}
    \delta W=-\frac{\sqrt{2}\GF}{2}\vec{\Delta}_\mu\cdot \partial_t\delta \vec{P}^\mu.
\end{equation}
This is the final expression for the work done per unit time on a test neutrino; therefore, by definition, $-\delta W$ is the energy given per unit time to the collective field, i.e., $\partial_t \int d^3 \br \mathcal{E}=-\int d^3 \br \delta W$, where $\mathcal{E}$ is the total energy of the field. Introducing finally the flavor isospin components 
\begin{equation}
    \vec{\Delta}^\mu=\left(\frac{\psi^\mu+\psi^{\mu,*}}{2},\frac{\psi^\mu-\psi^{\mu,*}}{2i},g^\mu\right)
\end{equation}
where $\psi^\mu$ will ultimately correspond to the flavomon field, while $g^\mu$ is the fluctuating part of the $z$ isospin component, and analogous definitions for $\delta \psi^\mu$ and $\delta g^\mu$ for $\delta \vec{P}^\mu$, we obtain
\begin{equation}
    \delta W=-\frac{\sqrt{2}\GF}{2}\left[g^\mu \partial_t \delta g_\mu+\frac{\psi^\mu \partial_t \delta \psi^*_\mu+\psi^*_\mu \partial_t \delta \psi^\mu}{2}\right].
\end{equation}

It only remains to express $\delta W$ in terms of field quantities alone, removing the test neutrino properties which were introduced only as a conceptual aid. To do so, we notice that by definition the total field $\psi^\mu$ must equal the sum of the test neutrino field and the medium field
\begin{equation}
    \psi^\mu=\delta \psi^\mu+\psi^\mu_{\rm med}.
\end{equation}
In turn, the medium polarization vector $\psi^\mu_{\rm med}$ in linear theory can be related to the total field via the flavor susceptibility
introduced in Ref.~\cite{Fiorillo:2024bzm}
\begin{equation}
    \psi^\mu_{\rm med}=\hat{\chi}^{\mu\nu}\psi_\nu.
\end{equation}
As in the plasma case discussed in Sect.~A, this expression is understood as a convolution of expressions in Fourier space, where ${\psi}^\mu_{\rm med}(\Omega,\bK)={\chi}^{\mu\nu}(\Omega,\bK) {\psi}_\nu(\Omega,\bK)$.

Introducing the dielectric tensor $\hat{\varepsilon}^{\mu\nu}=g^{\mu\nu}-\hat{\chi}^{\mu\nu}$, we have $\delta \psi^\mu=\hat{\varepsilon}^{\mu\nu}\psi_\nu$. For the fluctuating $z$ components $g^\mu$ and $\delta g^\mu$ a similar relation can be formally introduced, but to linear order the medium itself does not produce any response to an external $\delta g^\mu$ -- the perturbations in the $z$ component are quadratic, as well known. For our purpose, we therefore neglect the work done by these fluctuating fields, which must be derived at highest orders in the nonlinear theory. Hence, our final result is
\begin{equation}
    \delta W=-\frac{\sqrt{2}\GF}{4}\left[\psi^\mu \partial_t (\hat{\varepsilon}^*_{\mu\nu}\psi^{*,\nu})+\psi^*_\mu \partial_t(\hat{\varepsilon}^{\mu\nu}\psi_\nu)\right],
\end{equation}
allowing us first of all to find the energy dissipated per unit time by the field. Passing to the Fourier components we find that the total energy dissipated over the entire volume and time is 
\begin{widetext}
\begin{equation}\label{eq:derivative_field_energy}
    \partial_t E_{\rm field}=-\int d^3\br \delta W=\frac{\sqrt{2}\GF i}{4}\sum_\bK \int \frac{d\Omega}{2\pi}\int \frac{d\Omega'}{2\pi} {\psi}^\mu(\Omega,\bK){\psi}^{\nu,*}(\Omega',\bK)\left[-\Omega \varepsilon_{\mu\nu}(\Omega,\bK)+\Omega' \varepsilon^*_{\mu\nu}(\Omega',\bK)\right] e^{i(\Omega'-\Omega)t}.
\end{equation}
If we integrate over $t$ from $-\infty$ to $+\infty$, we finally find
\begin{equation}
    Q=\frac{\sqrt{2}\GF i}{4}\sum_\bK \int \frac{d\Omega}{2\pi} {\psi}^\mu(\Omega,\bK){\psi}^{\nu,*}(\Omega,\bK)\Omega\left[\varepsilon^*_{\mu\nu}(\Omega,\bK)-\varepsilon_{\mu\nu}(\Omega,\bK)\right].
\end{equation}
Hence, we recover the result, familiar from electrodynamics and already obtained in Ref.~\cite{Fiorillo:2024bzm}, that the imaginary part of the dielectric tensor describes the dissipation of the field energy. If the latter is not too large, we can approximately define an expression for the field energy itself. We thus return to Eq.~\eqref{eq:derivative_field_energy} and, before averaging over time, we integrate it in time. The final result, after performing the time average and neglecting the imaginary part of $\varepsilon^{\mu\nu}$, is
\begin{equation}\label{eq:energy_field}
    E_{\rm field}=\frac{\sqrt{2}\GF}{4}\sum_\bK \int \frac{d\Omega}{2\pi}{\psi}^\mu(\Omega,\bK){\psi}^{\nu,*}(\Omega,\bK)\partial_\Omega\left[\Omega \varepsilon_{\mu\nu}(\Omega,\bK)\right].
\end{equation}
\end{widetext}
This is our final expression for the field energy, which is well-defined only if the imaginary part of $\varepsilon^{\mu\nu}$ is small, so that there is no rapid energy exchange between field and neutrinos. This expression is fully equivalent to the energy of an electric field in a plasma \cite{landau2013electrodynamics}.

The field energy thus identified allows us to finally introduce a concrete definition of flavomons. The collective oscillations of the flavor field are defined by the equation $\varepsilon^{\mu\nu}(\Omega_\bK,\bK){\psi}_\nu(\Omega_\bK,\bK)=0$, where $\Omega_\bK$ is the eigenfrequency associated with the collective oscillation. As usual, we stick to the assumption of weak instability, in which $\Omega_\bK$ has only a small imaginary part. (Even then, there generally are Landau-damped modes for which $\mathrm{Im}(\Omega_\bK)\gg \mathrm{Re}(\Omega_\bK)$, but since they are damped, their strong damping simply means that they can be neglected altogether.) Hence, in an ensemble average, we can write
\begin{equation}
    \langle {\psi}^\mu(\Omega,\bK){\psi}^{\nu,*}(\Omega,\bK)\rangle=|{\psi}_{\bK}|^2 e^\mu_{\bK} e^\nu_{\bK}2\pi \delta(\Omega-\Omega_\bK),
\end{equation}
where $e^\mu_\bK$ is the eigenvector associated with the frequency $\Omega_\bK$. If for a given $\bK$, there are multiple branches of collective oscillations, this equation can be obviously generalized. Hence, the field energy can finally 
be written
\begin{equation}
    E_{\rm field}=\frac{\sqrt{2}\GF}{4}\sum_\bK |{\psi}_\bK|^2\Omega_\bK \mathcal{Z}_{\bK}^{-1},
\end{equation}
where
\begin{equation}
      \mathcal{Z}_{\bK}^{-1}=e^\mu_{\bK}e^{*\nu}_{\bK}\partial_\Omega \varepsilon_{\mu\nu}(\Omega,\bK)\big|_{\Omega=\Omega_\bK}.
\end{equation}
From here, we identify the effective number of flavomons
\begin{equation}
    N_\bK=\frac{\sqrt{2}\GF}{4}|{\psi}_\bK|^2\mathcal{Z}_{\bK}^{-1},
\end{equation}
such that $E_{\rm field}=\sum_\bK N_\bK \Omega_\bK$.

For a particle physicist, it is also useful to interpret the dielectric tensor in a slightly different way; since $\varepsilon^{\mu\nu}$ can be interpreted as the inverse flavomon propagator, its derivative $\mathcal{Z}_{\bK}^{-1}$ corresponds to the wavefunction renormalization factor, which is proportional to the derivative of the inverse propagator as usual.

We can also verify that this definition consistently leads to the conclusion that the flavomon carries one unit of electron number and minus one unit of muon number, similar to a gluon which carries one unit of color and one of anticolor. To do this, we return to Eq.~\eqref{eq:test_neutrino_polarization}, sum over $\bp$ and determine the total $e-\mu$ lepton-number dissipated $\delta g=\sum_\bp \delta g_\bp$ as
\begin{equation}
    \partial_t \delta g+\partial_i \delta g^i=\frac{\sqrt{2}\GF i}{2}\left(\psi^\mu \delta \psi^*_\mu-\psi^*_\mu \delta \psi^\mu\right).
\end{equation}
Going through the same steps as for the energy, we determine the lepton number in the field as (minus) the integral over time of the right-hand side with $\delta\psi_\mu=\hat{\varepsilon}_{\mu\nu}\psi^\nu$. Hence, the total $e-\mu$ lepton number $L_{\rm field}$ in the field is
\begin{widetext}
    \begin{equation}
       \nonumber L_{\rm field}=-\frac{\sqrt{2}\GF}{2}\sum_\bK \int \frac{d\Omega}{2\pi}\psi^\mu(\Omega,\bK) \psi^{\nu,*}(\Omega,\bK)\partial_\Omega \varepsilon_{\mu\nu}(\Omega,\bK)=-\sum_\bK 2 N_\bK. 
    \end{equation}
By comparing with Eq.~\eqref{eq:energy_field}, we immediately see that a flavomon carries two units of $e-\mu$ lepton number.

\section{C.~Systematic introduction of flavomons in a quantum field theory.}

The flavomon has been introduced in a somewhat heuristic way, by regarding the classical flavor field as a collection of elementary excitations. However, the appearance of quantized excitations can also be retrieved in a systematic approach by starting from the quantum field theory (QFT) of interacting neutrinos. We therefore start from the full Hamiltonian
\begin{equation}\label{eq:many_body_hamiltonian}
    H=\sum_\bp \varepsilon_\bp a^\dagger_{i,\bp}a_{i,\bp}+\frac{\sqrt{2}\GF}{2}\sum_{\bp,\bK,\bp'}a^\dagger_{i,\bp+\frac{\bK}{2}}a_{j,\bp-\frac{\bK}{2}}a^\dagger_{j,\bp'-\frac{\bK}{2}}a_{i,\bp'+\frac{\bK}{2}}v\cdot v',
\end{equation}    
\end{widetext}
which fully describes the system of interacting neutrinos, without any mean-field assumption. The only required approximation to obtain this many-body Hamiltonian is the weak-inhomogeneity assumption $|\bK|\ll|\bp|,|\bp'|$.

\begin{figure*}
    \includegraphics[width=0.5\textwidth]{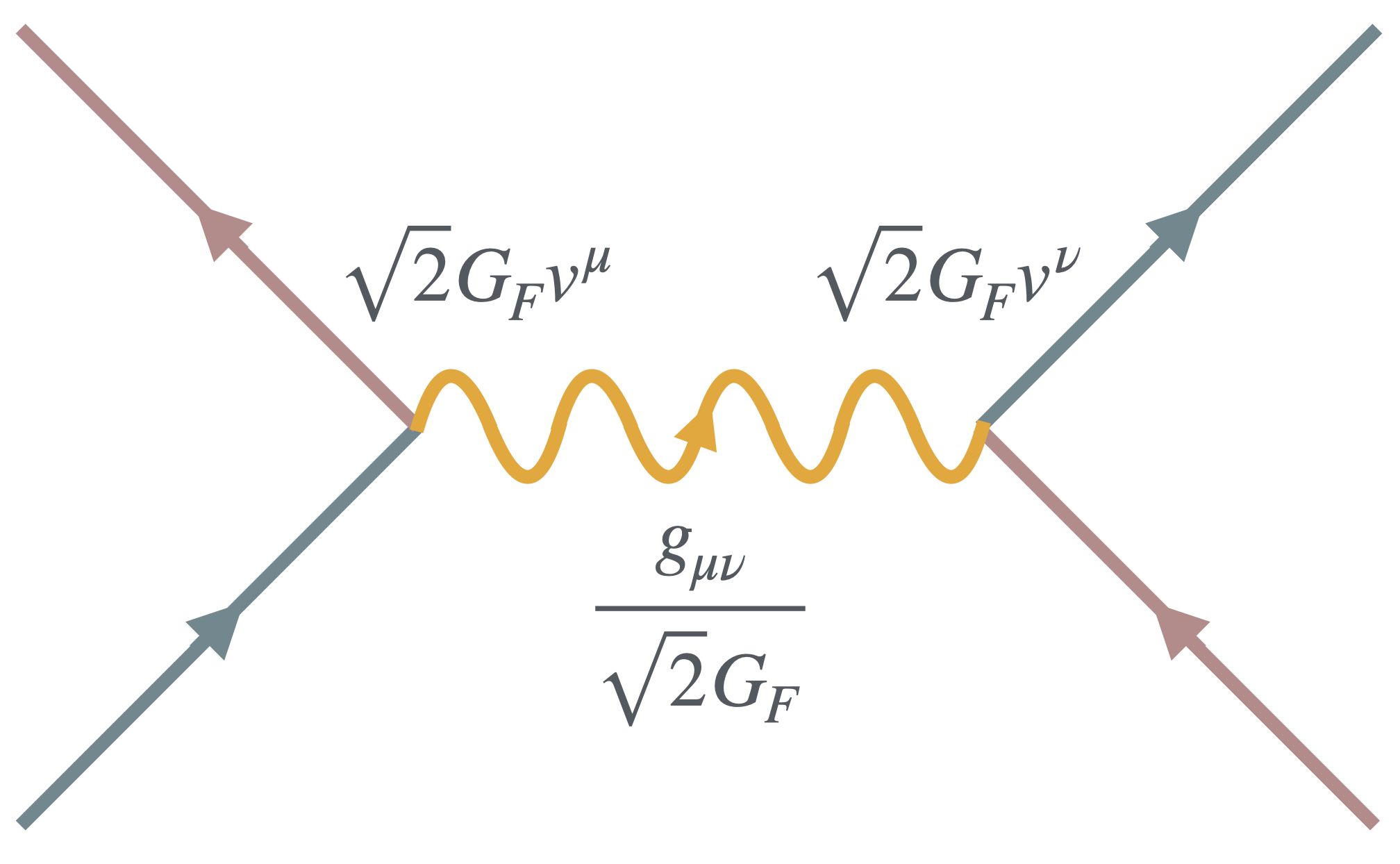}
    \includegraphics[width=\textwidth]{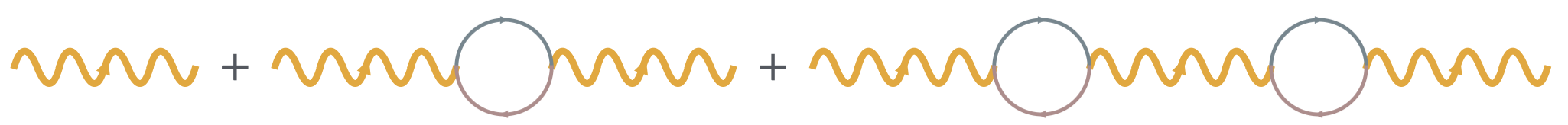}
    \caption{\textit{(Top)} Neutrino-neutrino interaction at tree level, mediated by a fictitious flavomon with a static, short-range propagator. \textit{(Bottom)} Diagrams contributing to the renormalization of the flavomon propagator in RPA.}\label{fig:RPA}
\end{figure*}

In QFT, we may describe the interaction terms diagrammatically by the tree-level diagram in Fig.~\ref{fig:RPA}. At this stage, the wavy line used to describe the interaction between neutrinos is nothing more than a symbol, and does not correspond to a dynamical degree of freedom, since the interaction between neutrinos is assumed to be local and instantaneous. Nevertheless, even at this stage, it is convenient to describe the interaction by means of an effective coupling $\sqrt{2}\GF v^\mu$ associated with each neutrino interaction vertex, and a propagator $D_{\mu\nu}=(\sqrt{2}\GF)^{-1}g_{\mu\nu}$ associated with the wavy line. One easily verifies that this prescription exactly reproduces the Feynman rules of the Hamiltonian in Eq.~\eqref{eq:many_body_hamiltonian}, and in fact may be obtained explicitly by a Hubbard-Stratonovich transformation. Since the propagator does not depend on any frequency or wavevector, the instantaneous and local nature of the interaction is clear.

\begin{figure*}
    \includegraphics[width=0.8\textwidth]{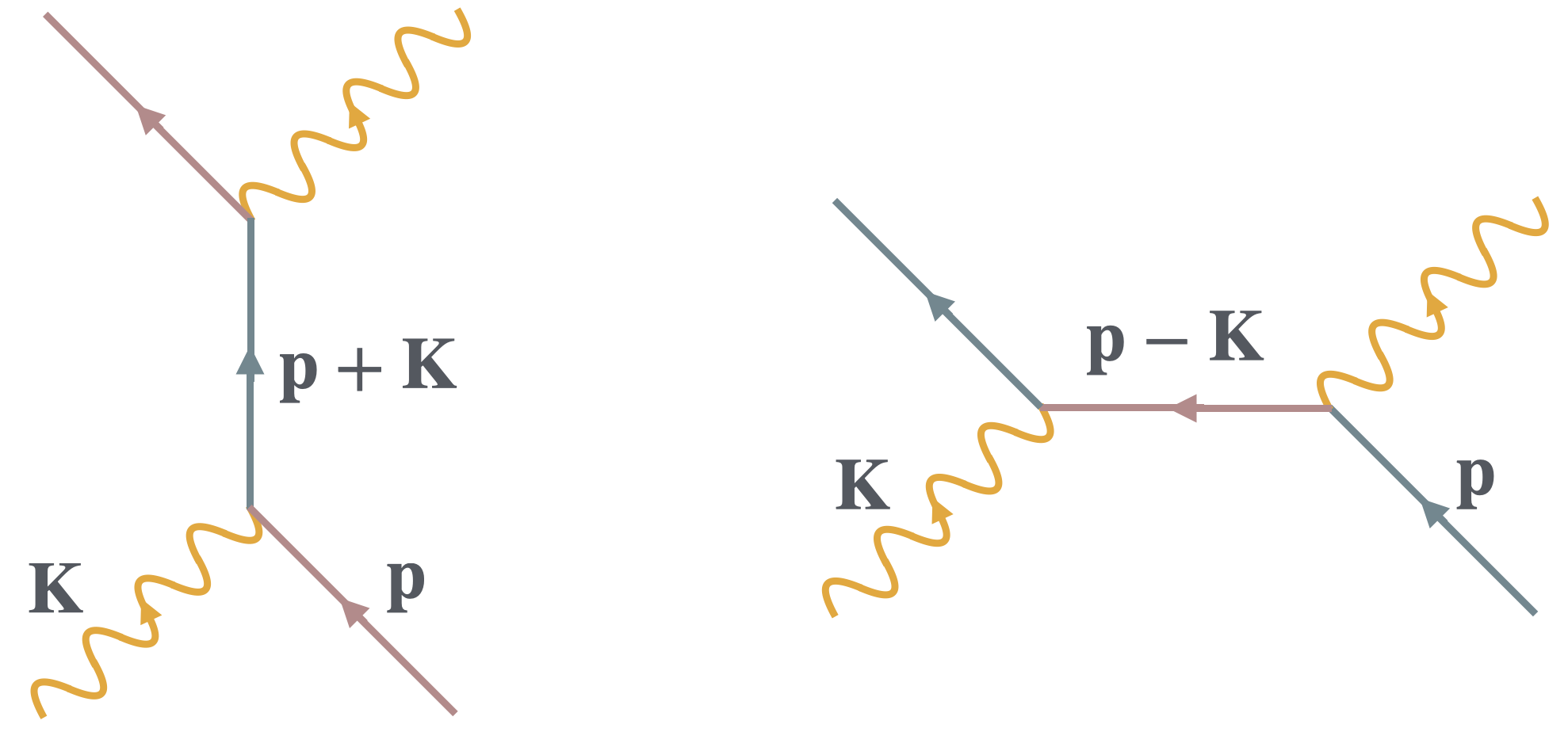}
    \caption{Processes contributing to the forward scattering of flavomons, which determines their self-energy in the neutrino medium. Time flows upward along the vertical direction.}\label{fig:fwscattering}
\end{figure*}

However, the situation changes once we go to higher orders in perturbation theory. In a medium, the dominant diagrams that renormalize the neutrino-neutrino interaction are the random phase approximation (RPA) diagrams, or bubble diagrams, that are shown in Fig.~\ref{fig:RPA}. The renormalized propagator may then be obtained from the recursion relation
\begin{equation}
    i\mathcal{D}_{\mu\nu}=iD_{\mu\nu}+(iD_{\mu\alpha})(-i\Sigma_{\alpha\beta})(i\mathcal{D}_{\beta\nu}).
\end{equation}
The self-energy part $\Sigma_{\alpha\beta}$, corresponding to the $\nu_e$ and $\nu_\mu$ loop, is therefore the crucial part of this calculation. To explicitly obtain it in a systematic way requires the use of nonequilibrium field theory, since the neutrino medium is not in flavor equilibrium. However, to lowest order the self-energy can be obtained from the amplitude for forward scattering of the excitations described by the wavy lines, which we will already refer to as flavomons even though at this stage of the calculation they do not yet appear as dynamical excitations. This is indeed the same result that is used to obtain the self-energy of the plasmon from the forward scattering of a photon off an electron, a calculation that we have recently reviewed in the appendix of Ref.~\cite{Fiorillo:2024bzm}.

Flavomon forward scattering
is described by the two diagrams in Fig.~\ref{fig:fwscattering}. The momenta associated with each line are labeled in the figure; the energy of a $\nu_\mu$ with momentum $\bp$ is $p^0=|\bp|+\sqrt{2}\GF n_\mu$, while the energy of a $\nu_e$ with the same momentum is $p^0=|\bp|+\sqrt{2}\GF n_e$, as discussed in the main text. Finally, the frequency of the flavomon line is denoted by $\Omega$, and is for now an arbitrary variable. We may therefore obtain its amplitude from the expression
\begin{widetext}
\begin{equation}
    \Sigma_{\alpha\beta}=-\int \frac{d^3\bp}{(2\pi)^3}2 \GF^2v_\alpha v_\beta \left[\frac{n_{\mu,\bp}}{|\bp|+\sqrt{2}\GF G\cdot v+\Omega-|\bp+\bK|}+\frac{n_{e,\bp}}{|\bp|-\sqrt{2}\GF G\cdot v-\Omega-|\bp-\bK|}\right].
\end{equation}    
\end{widetext}
Expanding for small $|\bK|$ in the denominator, this result finally gives
\begin{equation}
    \Sigma_{\alpha\beta}=\int \frac{d^3\bp}{(2\pi)^3}2\GF^2  \frac{G_\bp v_\alpha v_\beta}{k\cdot v+i\varepsilon}=\sqrt{2}\GF \chi_{\alpha\beta},
\end{equation}
where we have recognized the appearance of the flavor susceptibility $\chi_{\alpha\beta}$, and we have added an $i\epsilon$ regularizer to ensure the retarded nature of the propagator. Therefore, the resummed retarded propagator obeys
\begin{equation}
    (D_{\mu\alpha}^{-1}-\Sigma_{\mu\alpha})\mathcal{D}^{\alpha\nu}=\delta_\mu^\nu,
\end{equation}
which is easily seen to be identical to
\begin{equation}
    \varepsilon_{\mu\alpha}\mathcal{D}^{\alpha\nu}=D_\mu^\nu.
\end{equation}
Thus, differently from the tree-level propagator, the resummed one has been endowed with dynamics by the neutrino medium, and, due to the RPA resummation, has acquired a life of its own, so to speak. In particular, the poles of the propagator are determined by the eigenvalue equation
\begin{equation}
    \varepsilon_{\mu\nu}(\Omega_{\bK},\bK)e^\nu_{\bK}=0,
\end{equation}
and around these poles, the propagator can be expanded as
\begin{equation}
    \mathcal{D}^{\mu\nu}\simeq e^\mu_{\bK} e^\nu_{\bK}\frac{\mathcal{Z}_{\bK}}{\Omega-\Omega_\bK}.
\end{equation}
Therefore, close to such poles, the flavomon propagator has the same behavior as if there was an elementary particle with energy $\Omega_\bK$ and a wavefunction renormalized by a factor $\sqrt{\mathcal{Z}_{\bK}}$. 

This interpretation also helps to clarify a conceptual point; even though the neutrino-neutrino interaction is short ranged and instantaneous, neutrinos can develop dynamical collective excitations. Our new derivation for their dispersion relation makes it clear that the dynamics does not come from the interaction, but rather from the continuous forward scattering on medium neutrinos through the bubble diagrams. So there is no fundamental difference from the case of the plasmon, even though the electromagnetic interaction is intrinsically long ranged
in our case, it is the medium that endows the flavor interactions with dynamics and range. Borrowing from the condensed-matter language, the flavomon is a bound state of a particle with electron flavor and a hole with muon flavor.

We should also clarify that the original theory we started with is a complete many-body theory. Our resummation of the RPA diagrams, and the subsequent treatment of the flavomon as an independent excitation that interacts to lowest order in $\GF$, completely reproduces the effects of mean field theory. The diagrammatic approach, however, also allows us to see graphically what is missed by this approximation. In the chosen terminology adopted here, only on-shell flavomons produced by neutrinos via Cherenkov radiation is included; this process has a probability proportional to $\GF$ and therefore can happen sufficiently rapidly. Going beyond mean field theory requires to include diagrams in which flavomons can also be off-shell, such as the diagram in Fig.~\ref{fig:RPA} (top). It is clear that this diagram, with an off-shell flavomon, describes neutrino-neutrino scattering, which does lead to flavor exchange mediated by an off-shell flavomon, but only over long timescales, since the probability is proportional to $\GF^2$. This immediately shows that over the astrophysically relevant timescales, only the RPA-resummed diagrams will relevantly contribute. This is indeed the same situation as in plasmas or in many-body systems (e.g., magnetic systems, Fermi liquids), in which the RPA approximation is usually adopted.

\begin{figure*}
    \includegraphics[width=\textwidth]{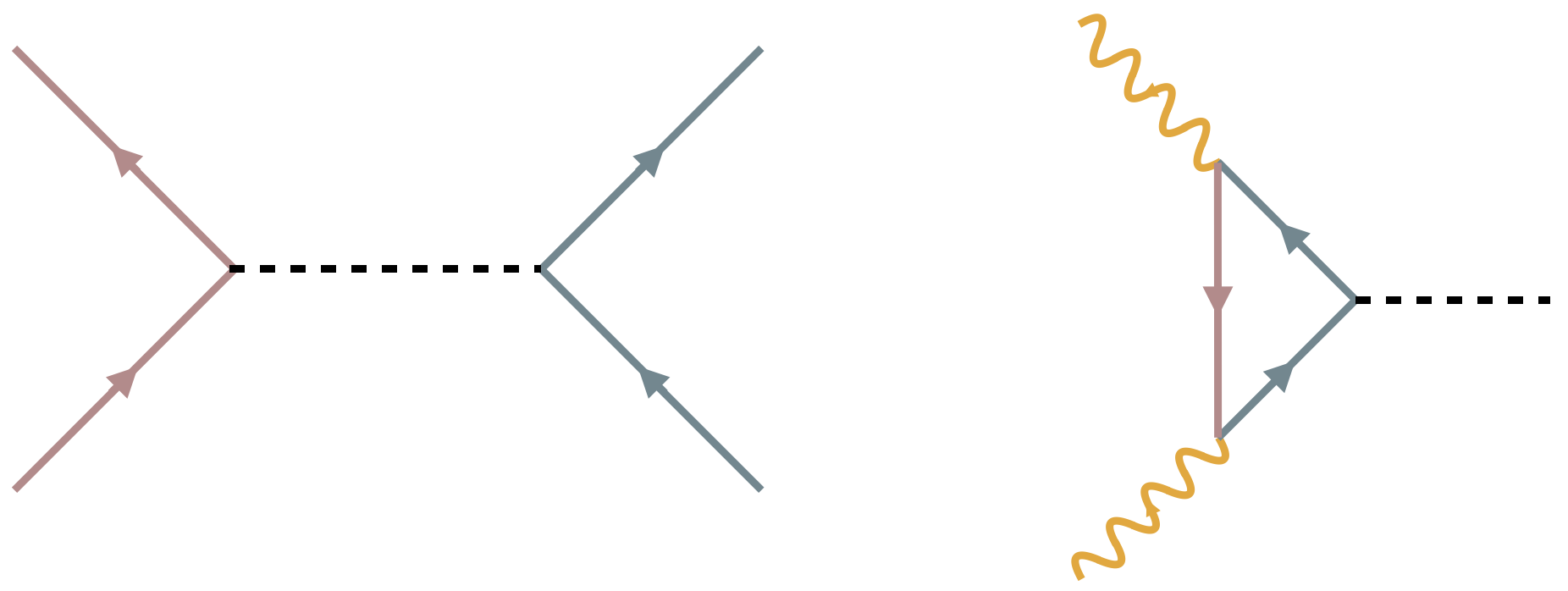}
    \caption{\textit{(Left)} Neutrino-plasmon as the mediator of flavor-diagonal interactions. For graphical reasons, we only show the interaction between an electron and a muon neutrino, but the same interaction also couples electron and muon neutrinos among themselves. \textit{(Right)} Flavomon-neutrino-plasmon interaction mediated by the triangle neutrino diagram; this interaction is responsible for neutrino-plasmon production in the non-linear stage of instability.}\label{fig:neutrinoplasmon}
\end{figure*}

Our systematic introduction of flavomons also allows us to clarify an additional point which is not completely transparent in the formalism introduced in the main text. The number density of flavomons $N_{\bK}$ does not change on small scales, which are instead associated only with the wavelength of the flavomon itself. From the kinetic equation for the neutrino number density, one would therefore conclude that also $n_{e,\bp}$ and $n_{\mu,\bp}$ only change on large length scales. This seems in immediate contradiction with the standard picture of collective flavor conversions, which leads to small-scale changes in the number density of neutrinos with different flavors. The resolution of the contradiction is that we have implicitly performed a separation of scales in defining $n_{e,\bp}$ and $n_{\mu,\bp}$, so that only the large-scale structures are included in their definition. In other words, the neutrino number density used in the main text is
\begin{equation}
    n_{i,\bp}=\sum_{|\bq|\ll|\bK|} \langle a^\dagger_{i,\bp-\frac{\bK}{2}} a_{i,\bp+\frac{\bK}{2}}\rangle e^{i\bK\cdot \br}.
\end{equation}
This separation of scales is implicit when we compute the decay rate, as we did in the main text, by treating neutrinos and flavomons as particles, since this intrinsically requires that their distributions change over distances much larger than their wavelength, both of the neutrino and the flavomon.

The small-scale density fluctuations can therefore be associated with a separate kind of elementary excitation, closer in spirit to the standard plasmon, since it corresponds to a $\nu_e$ and $\nu_\mu$ density fluctuation. These excitations appear from the terms with $i=j$ in the many-body Hamiltonian Eq.~\eqref{eq:many_body_hamiltonian}, which we rewrite as
\begin{equation}
    H_{\rm plasmon}=\frac{\sqrt{2}\GF}{4}\sum_{\bp,\bK,\bp'}\left[g^\mu_{\bK}g_{\mu,-\bK}+n^\mu_{\bK} n_{\mu,-\bK}\right],
\end{equation}
where we introduce the operators
\begin{eqnarray}
    g^\mu_{\bK}=(a^\dagger_{e,\bp-\frac{\bK}{2}}a_{e,\bp+\frac{\bK}{2}}-a^\dagger_{\mu,\bp-\frac{\bK}{2}}a_{\mu,\bp+\frac{\bK}{2}})v^\mu,\\
    n^\mu_{\bK}=(a^\dagger_{e,\bp-\frac{\bK}{2}}a_{e,\bp+\frac{\bK}{2}}+a^\dagger_{\mu,\bp-\frac{\bK}{2}}a_{\mu,\bp+\frac{\bK}{2}})v^\mu,
\end{eqnarray}
where only the short-scale variations corresponding to $|\bK|$ comparable with the flavomon wavelength are kept. Here $g^\mu_{\bK}$ is an operator corresponding to a density fluctuation with opposite sign for $\nu_e$ and $\nu_\mu$, whereas $n^\mu_\bK$ is a density fluctuation of the overall neutrino gas; the latter is generally negligible, as it does not directly couple to flavomons and therefore is not produced in the development of the instability.

The excitations associated with the field $g^\mu_\bK$ can now be introduced diagrammatically, as we did for the flavomons, by decomposing the interaction term via a broken line as in Fig.~\ref{fig:neutrinoplasmon} (left).  The role of these neutrino-plasmons during the early stages of instability development is inessential, since they are not generated at the linear level, and therefore are irrelevant for weak instabilities. However, if the number of flavomons in the medium becomes sufficiently large, the primary form of non-linear interaction for them is the generation of neutrino-plasmons via the triangle diagram shown in Fig.~\ref{fig:neutrinoplasmon} (right). Therefore, for the development of a higher-order perturbation theory of neutrinos and flavomons, introducing the neutrino-plasmons would be essential. Our scope in this work is more limited, so we only introduce them as an indication of the way to follow to systematically develop this theory to higher orders.

\section{D.~Flavomon dispersion for a specific example.}
\label{app:example}

\begin{figure}
    \centering
    \includegraphics[width=1.0\columnwidth]{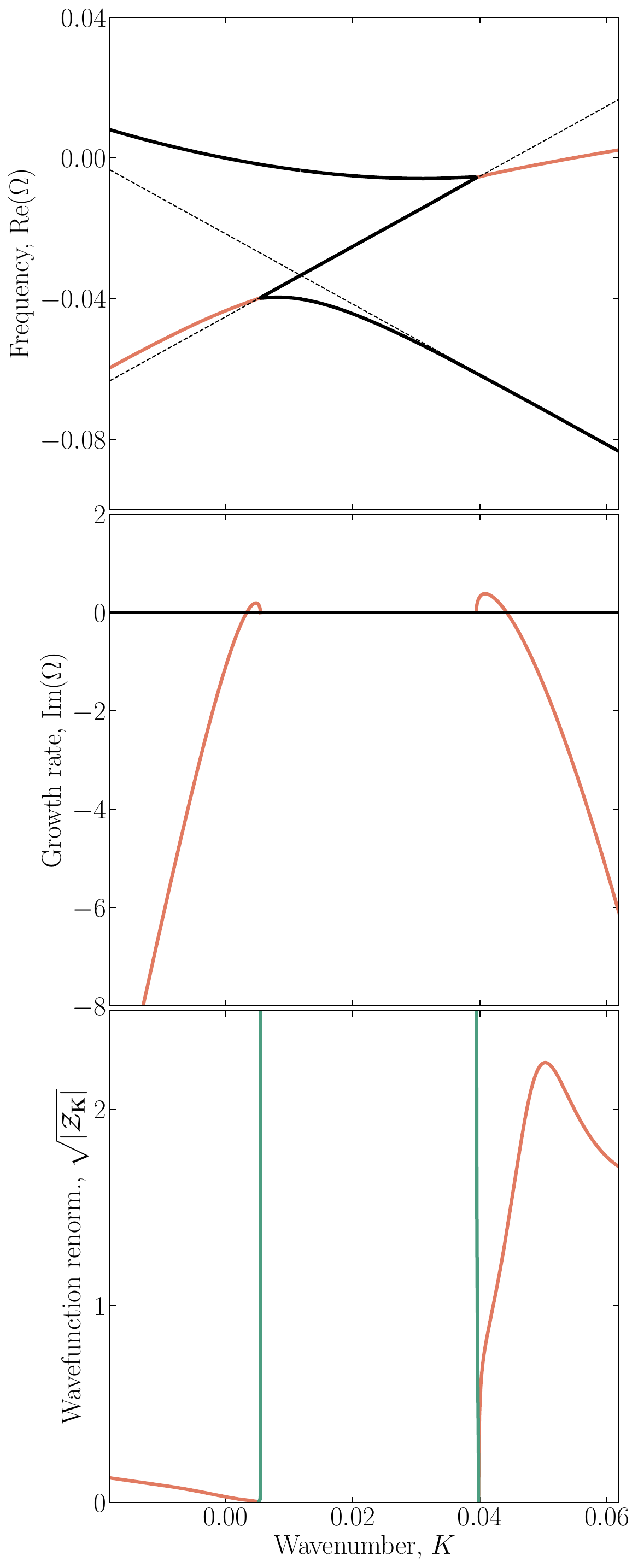}
    \caption{Dispersion relation for longitudinal flavomons in the axially symmetric neutrino angular distribution of Eq.~\eqref{eq:G-distribution}. We show real modes in black and unstable or Landau-damped modes in red. In the top panel, the light cone is shown with thin dashed lines. In the bottom panel, we show in green the lines corresponding to the region where $\mathcal{Z}_\bK<0$.}
    \label{fig:example}
\end{figure}

In this section, we provide a specific example of the dispersion and properties of longitudinal flavomons in a concrete case. We choose an angular distribution that has a very shallow crossing, so that the resulting instability is very weak, as one expects in the early stages of an instability; our example is the same that we recently used in Ref.~\cite{Fiorillo:2025ank}, with
\begin{equation}\label{eq:G-distribution}
    \sqrt{2}\GF G(v)=0.03-0.0302 \exp\left[-(1-v)^2\right];
\end{equation}
here $G(v)$ is the axially symmetric velocity distribution for the electron-minus-muon lepton number
\begin{equation}
    G(v)=\int_0^{\infty}\frac{p^2 dp}{4\pi^2}(n_{e,\bp}-n_{\mu,\bp}).
\end{equation}
The units of space and time are chosen such that the energy scale is dimensionless.

To obtain the dispersion relation of the longitudinal plasmons, we use the dielectric tensor for $\bK$ directed along the axis of symmetry~\cite{Fiorillo:2024uki}
\begin{equation}
    \varepsilon^{\mu\nu}=\begin{pmatrix}
        1-I_0 & 0 & 0 & -I_1\\
        0 & -1-\frac{I_0-I_2}{2} & 0 & 0 \\
        0 & 0 & -1-\frac{I_0-I_2}{2} & 0 \\
        -I_1 & 0 & 0 & -1-I_2
    \end{pmatrix},
\end{equation}
where 
\begin{equation}
    I_n=\int_{-1}^{+1}\frac{G(v) v^n}{\omega-k v+i\epsilon}dv.
\end{equation}
This expression leads to the dispersion relation for longitudinal flavomons
\begin{equation}
    (1-I_0)(1+I_2)+I_1^2=0,
\end{equation}
and the polarization vector for a given eigenmode must satisfy $\varepsilon^{\mu\nu}e_\nu=0$, so it can be chosen as
\begin{equation}
    e_0=I_1
    \quad\hbox{and}\quad
    e_1=1-I_0;
\end{equation}
the overall normalization of the polarization vector is of course arbitrary. With this choice, the wavefunction renormalization is
\begin{eqnarray}
    \mathcal{Z}_\bK&=&\frac{1}{\partial_\Omega \varepsilon^{\mu\nu} e_{\mu,\bK} e^*_{\nu,\bK}}
    \\ \nonumber
    &=&-\frac{1}{I_0' |I_1|^2+2I_1'\mathrm{Re}\left[I_1^*(1-I_0)\right]+I_2'|1-I_0|^2},
\end{eqnarray}
where $I_n'=\partial_\Omega I_n=\partial_\omega I_n$.

Figure~\ref{fig:example} shows the dispersion relation for the eigenmodes of the system, and therefore for the flavomons. These coincide with the ones shown in Ref.~\cite{Fiorillo:2025ank} for the fast-unstable case; the real branches have a characteristic S-shaped structure, which produces the unstable bands immediately outside of the light cone. When these bands cross the light cone, they remain unstable for a narrow interval of wavenumber, corresponding to flavomons resonantly interacting with the neutrinos in the ``flipped'' region of the crossing, and then turn to Landau-damped modes once they start resonating with neutrinos in the electron-neutrino-dominated angular region. In the flavomon language, this is because they are now interacting with a region primarily populated with electron neutrinos, and therefore they are absorbed by $\nu_e\psi\to\nu_\mu$ more rapidly than they are produced by $\nu_\mu\to\nu_e\psi$.

In the bottom panel of Fig.~\ref{fig:example}, we also show the wavefunction renormalization $\mathcal{Z}_\bK$ for the unstable and Landau-damped modes, which are the most interesting ones since they interact with neutrinos resonantly. While we directly show $\sqrt{|\mathcal{Z}_\bK|}$, we find that for both the left and the right branches $\mathcal{Z}_\bK$ is positive (corresponding to flavomons) until it touches $\mathcal{Z}_\bK=0$, which is visible for both branches in Fig.~\ref{fig:example}. This happens when the mode turns from subluminal to superluminal, since at this point the dielectric tensor becomes singular and therefore its derivatives diverge. Once the mode turns superluminal, the wavefunction renormalization becomes negative (we have superluminal antiflavomons) and diverges to infinity as the mode approaches the branching point where it becomes real; this is because at the branching point we have by definition that $\partial_\Omega \varepsilon^{\mu\nu} e_\mu e^*_\nu=0$~\cite{Fiorillo:2024uki}. Numerically, we are of course unable to track precisely the nature of the solution in the very narrow interval of superluminal unstable modes, so the curves are interrupted in Fig.~\ref{fig:example}. These superluminal modes have a very large wavefunction renormalization, but they cannot be produced on-shell by neutrinos because they are superluminal; in reality, they can be produced due to the resonance broadening with a width $\mathrm{Im}(\Omega_\bK)$, which tends to $0$ as $|\mathcal{Z}_{\bK}|\to \infty$, so there is no inconsistency in this infinite wavefunction renormalization.

Our general lesson from this special example is that in a medium that is dominated by electron neutrinos, the vast majority of flavomon states do not include antiflavomons. This is in fact a consequence of lepton number conservation: as proven in Refs.~\cite{Johns:2024bob,Fiorillo:2024fnl}, this implies that in the absence of an angular crossing there cannot be any fast instability. Therefore, in a medium that is dominated along any direction by electron neutrinos, there cannot be antiflavomon states, which would otherwise lead to an instability via $\nu_e\to\nu_\mu\overline{\psi}$.

\section{E.~Flavomons in a medium in equilibrium.}

We have introduced the concept of flavomons in an unstable medium, as this is the context in which they are practically useful. However, exactly as plasmons, flavomons should also appear in equilibrium, provided that the medium is endowed with a non-vanishing $e-\mu$ lepton number. This situation is particularly useful to test fundamental properties of the concept of flavomons, such as their intrinsic stability.

A medium in equilibrium must have isotropic thermal $\nu_e$ and $\nu_\mu$ distributions, with generically different chemical potentials $\mu_e$ and $\mu_\mu$, where we assume $\mu_e>\mu_\mu$ without loss of generality. Without interactions with a background medium, the total $e$ and $\mu$ lepton numbers are separately conserved, giving rise to two separate chemical potentials. In this medium, the properties of flavomons in the fast regime are remarkably simple. The kinematic condition for $\nu_\mu\to\nu_e\psi$ is independent of the neutrino energy, and so only the energy-integrated distributions are required. Therefore, any isotropic $\nu_e$ and $\nu_\mu$ distribution with different number densities is a general proxy for a medium in thermal equilibrium.

\begin{figure}
    \centering
    \includegraphics[width=1.0\columnwidth]{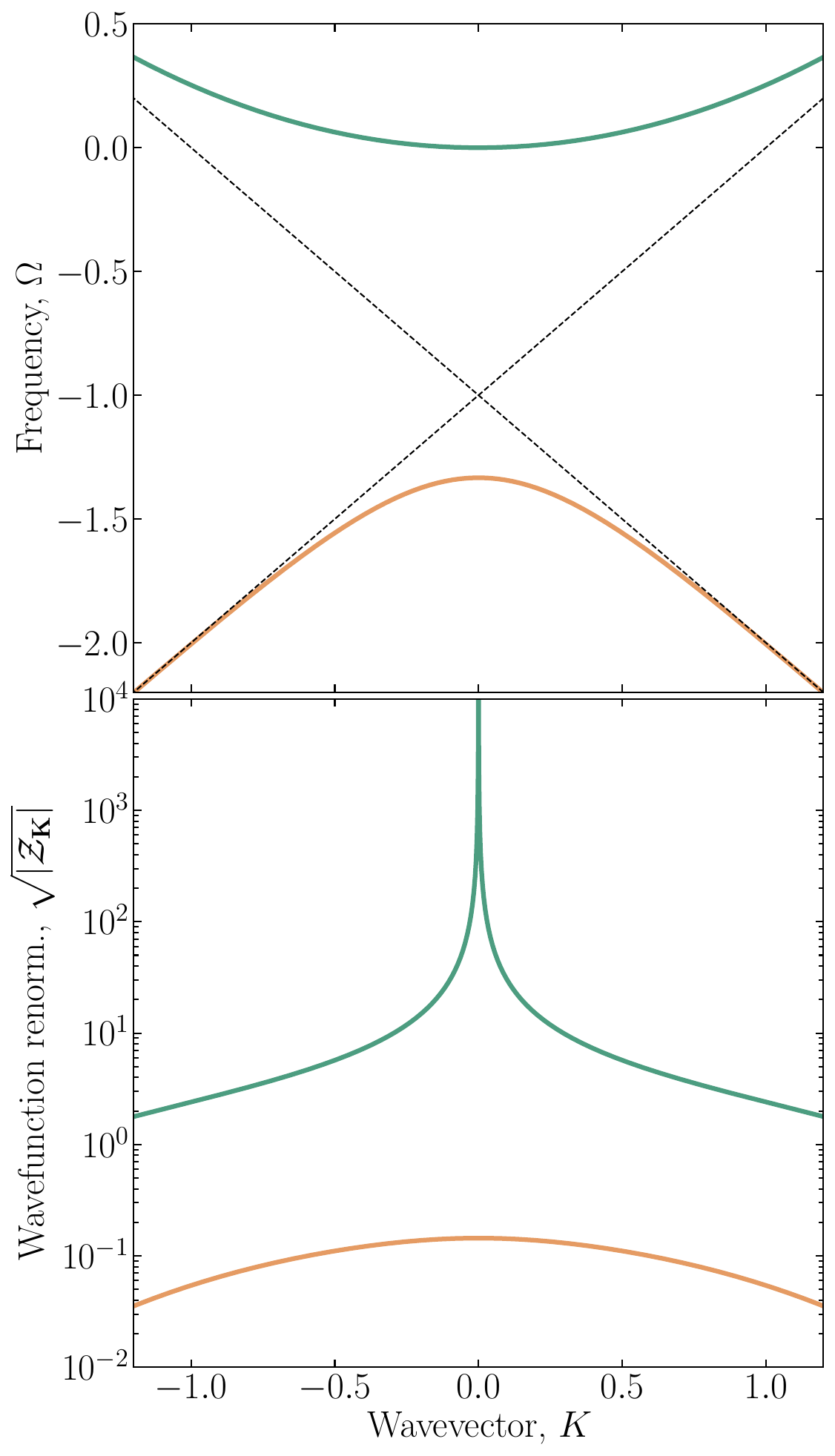}
    \vskip-6pt
    \caption{Dispersion relation and wavefunction renormalization for the longitudinal flavomons in a thermal and isotropic $\nu_e$--$\nu_\mu$--medium. We show in different colors the two branches, both of which have vanishing $\mathrm{Im}(\Omega)$. In the top panel, the light cone is shown with thin dashed lines.}
    \label{fig:equilibrium}
    \vskip-6pt
\end{figure}

Since the distributions are isotropic, we can use the same dispersion relation as in Sec.~D for the axially symmetric case. For the resulting longitudinal flavomons it is shown in Fig.~\ref{fig:equilibrium}. There exist only two branches of real superluminal modes, both have $\mathcal{Z}_\bK>0$, and so both of them describe flavomons rather than anti-flavomons. This agrees with our general finding that a $\nu_e$ dominated medium exhibits primarily flavomons. Importantly, however, the lower branch has negative frequencies, and therefore the field itself has negative energies.

The notion of negative energy in an equilibrium state is at first concerning because negative energies are associated with an intrinsic instability of the system. Formally, the Bose-Einstein distribution $n(E)=1/(e^{E/T}-1)$ turns negative when $E<0$, signaling the absence of a stable state. However, if the species is endowed with a chemical potential, the distribution is
\begin{equation}\label{eq:bose_distribution}
    n(E)=\frac{1}{e^{(E-\mu_\psi)/T}-1},
\end{equation}
and therefore a pathological behavior is expected only if $E<\mu_\psi$. For a neutrino-flavomon medium in thermal equilibrium, the flavomon chemical potential is fixed by the condition of chemical equilibrium $\mu_\psi=\mu_\mu-\mu_e<0$. 

The flavomon states have very small energies, being determined by the refractive energy scale, compared to neutrinos with energies of the order of the chemical potentials. Likewise, $|\bK|\ll |\bp|$, reflecting the separation of scales between the neutrino Compton scale and the scale of flavor conversion. Therefore, in module we necessarily have $|\mu_\psi|\gg |\Omega|$. 
In a realistic environment, $|\mu_\psi|\sim 10$~MeV, whereas $|\Omega|\sim \sqrt{2}\GF(n_\nu-\overline{n}_\nu)\sim 10^{-8}$~MeV.
Indeed, that neutrinos and flavor waves exchange any energy had gone entirely unnoticed until recently \cite{Fiorillo:2024fnl}. As the flavomon energy is negligible compared with the negative chemical potential, no negative occupation numbers arise and there is no inconsistency.

Figure~\ref{fig:equilibrium} would suggest that at arbitrarily large $K$, the energy $|\Omega|$ would become arbitrarily large as well, eventually reaching $|\mu_\psi|$. However, in such a regime, the quantum kinetic equations would no longer apply. Moreover, we can also see from Fig.~\ref{fig:equilibrium}, as $K$ grows larger the frequency becomes closer and closer to the light cone, and in turn the wavefunction renormalization, which multiplies the coupling of the flavomons to the neutrinos, becomes ever smaller. Therefore, flavomons decouple from neutrinos and will not reach thermal equilibrium.

\section{F.~Comparison with miscidynamics.}

The discussion of the neutrino-flavomon system in equilibrium provides an opportunity to comment on the relation of our theory to the conjecture that the neutrino system, driven by flavor conversions, might reach thermal and chemical equilibrium in a framework termed miscidynamics \cite{Johns:2023jjt, Johns:2024dbe, Johns:2024bob}. In the language of flavomons, when an instability occurs, flavomons are produced and finally settle into a stationary state, but such a state is generally non-thermal. From the kinetic equations Eqs.~(17) and~(21) in the main text, it is clear that, while thermal and chemical equilibrium is certain to be a solution, it is not the one that is generally reached. 

We also stress that our kinetic equations for neutrinos and flavomons has
an entropy that can only increase with time; it does not coincide with the entropy hypothesized in the miscidynamics framework~\cite{Johns:2023jjt}, but rather is the sum of the neutrino and flavomon entropies. The irreversible dynamics which leads to the growth of this entropy is initially surprising, since it arises in a collisionless system, and is ultimately caused by the random phase approximation for the flavomon field. This is not a novel phenomenon, since also in collisionless plasma instabilities, the entropy of the electron-plasmon system grows
\cite{schKT}. Still, it does not generally grow to its maximum, which would correspond to thermal equilibrium.

The smallness of flavomon energies and momenta compared to the neutrino ones means that the latter cannot reach thermal and chemical equilibrium through flavomon emission and absorption -- this is the flavomon version of our earlier argument \cite{Fiorillo:2024fnl} that the kinetic energy reservoir is very weakly coupled to the refractive one, and therefore full equilibration between the two sectors cannot be achieved. Therefore, while our kinetic equations are naturally coarse-grained in the same sense as the coarse-grained density matrix in the miscidynamics framework, and as generally done in turbulent settings, we conclude that equilibrium in the thermodynamical sense is unlikely to be achieved.

Instead, what happens is that for any given momentum $\bp$ along which an instability ensues, $\nu_e$ and $\nu_\mu$ reach equilibrium (removal of the angular crossing), with the flavomon population growing until the condition for an instability is removed. This is the same picture as for the bump-on-tail instability of plasma physics \cite{schKT}, in which a bump of high-energy electrons can decay to lower energies producing plasmons, until the energy distribution flattens and the produced plasmons freeze after this plateau appears. If the produced plasmon number is sufficiently high, they can non-linearly interact with each other and reach a new form of equilibrium, which is also generally non-thermal but instead turbulent. We do not investigate here the final flavomon distribution that is reached after the development of the instability and the possible non-linear flavomon interactions.

\onecolumngrid


%% file: Draft-v2.bbl
\providecommand{\href}[2]{#2}\begingroup\raggedright\begin{thebibliography}{10}

\bibitem{pines1956collective}
D.~Pines, \emph{Collective energy losses in solids}, \href{https://doi.org/10.1103/RevModPhys.28.184}{\emph{Rev. Mod. Phys.} {\bfseries 28} (1956) 184}.

\bibitem{Pantaleone:1992eq}
J.~T. Pantaleone, \emph{{Neutrino oscillations at high densities}}, \href{https://doi.org/10.1016/0370-2693(92)91887-F}{\emph{Phys. Lett. B} {\bfseries 287} (1992) 128}.

\bibitem{Samuel:1993uw}
S.~Samuel, \emph{{Neutrino oscillations in dense neutrino gases}}, \href{https://doi.org/10.1103/PhysRevD.48.1462}{\emph{Phys. Rev. D} {\bfseries 48} (1993) 1462}.

\bibitem{Samuel:1995ri}
S.~Samuel, \emph{{Bimodal coherence in dense selfinteracting neutrino gases}}, \href{https://doi.org/10.1103/PhysRevD.53.5382}{\emph{Phys. Rev. D} {\bfseries 53} (1996) 5382} [\href{https://arxiv.org/abs/hep-ph/9604341}{{\ttfamily hep-ph/9604341}}].

\bibitem{Duan:2006an}
H.~Duan, G.~M. Fuller, J.~Carlson and Y.-Z. Qian, \emph{{Simulation of coherent nonlinear neutrino flavor transformation in the supernova environment: Correlated neutrino trajectories}}, \href{https://doi.org/10.1103/PhysRevD.74.105014}{\emph{Phys. Rev. D} {\bfseries 74} (2006) 105014} [\href{https://arxiv.org/abs/astro-ph/0606616}{{\ttfamily astro-ph/0606616}}].

\bibitem{Sawyer:2004ai}
R.~F. Sawyer, \emph{{``Classical'' instabilities and ``quantum'' speed-up in the evolution of neutrino clouds}},  \href{https://arxiv.org/abs/hep-ph/0408265}{{\ttfamily hep-ph/0408265}}.

\bibitem{Izaguirre:2016gsx}
I.~Izaguirre, G.~Raffelt and I.~Tamborra, \emph{{Fast Pairwise Conversion of Supernova Neutrinos: A Dispersion-Relation Approach}}, \href{https://doi.org/10.1103/PhysRevLett.118.021101}{\emph{Phys. Rev. Lett.} {\bfseries 118} (2017) 021101} [\href{https://arxiv.org/abs/1610.01612}{{\ttfamily 1610.01612}}].

\bibitem{Mirizzi:2015eza}
A.~Mirizzi, I.~Tamborra, H.-T. Janka, N.~Saviano, K.~Scholberg, R.~Bollig, L.~H{\"u}depohl and S.~Chakraborty, \emph{{Supernova Neutrinos: Production, Oscillations and Detection}}, \href{https://doi.org/10.1393/ncr/i2016-10120-8}{\emph{Riv. Nuovo Cim.} {\bfseries 39} (2016) 1} [\href{https://arxiv.org/abs/1508.00785}{{\ttfamily 1508.00785}}].

\bibitem{Tamborra:2020cul}
I.~Tamborra and S.~Shalgar, \emph{{New Developments in Flavor Evolution of a Dense Neutrino Gas}}, \href{https://doi.org/10.1146/annurev-nucl-102920-050505}{\emph{Ann. Rev. Nucl. Part. Sci.} {\bfseries 71} (2021) 165} [\href{https://arxiv.org/abs/2011.01948}{{\ttfamily 2011.01948}}].

\bibitem{Richers:2022zug}
S.~Richers and M.~Sen, \emph{{Fast Flavor Transformations}}, \href{https://doi.org/10.1007/978-981-15-8818-1_125-1}{\emph{Handbook of Nuclear Physics} (2022) 1} [\href{https://arxiv.org/abs/2207.03561}{{\ttfamily 2207.03561}}].

\bibitem{Volpe:2023met}
M.~C. Volpe, \emph{{Neutrinos from dense environments: Flavor mechanisms, theoretical approaches, observations, and new directions}}, \href{https://doi.org/10.1103/RevModPhys.96.025004}{\emph{Rev. Mod. Phys.} {\bfseries 96} (2024) 025004} [\href{https://arxiv.org/abs/2301.11814}{{\ttfamily 2301.11814}}].

\bibitem{Duan:2010af}
H.~Duan, A.~Friedland, G.~McLaughlin and R.~Surman, \emph{{The influence of collective neutrino oscillations on a supernova r-process}}, \href{https://doi.org/10.1088/0954-3899/38/3/035201}{\emph{J. Phys. G} {\bfseries 38} (2011) 035201} [\href{https://arxiv.org/abs/1012.0532}{{\ttfamily 1012.0532}}].

\bibitem{Wu:2017drk}
M.-R. Wu, I.~Tamborra, O.~Just and H.-T. Janka, \emph{{Imprints of neutrino-pair flavor conversions on nucleosynthesis in ejecta from neutron-star merger remnants}}, \href{https://doi.org/10.1103/PhysRevD.96.123015}{\emph{Phys. Rev. D} {\bfseries 96} (2017) 123015} [\href{https://arxiv.org/abs/1711.00477}{{\ttfamily 1711.00477}}].

\bibitem{Li:2021vqj}
X.~Li and D.~M. Siegel, \emph{{Neutrino Fast Flavor Conversions in Neutron-Star Postmerger Accretion Disks}}, \href{https://doi.org/10.1103/PhysRevLett.126.251101}{\emph{Phys. Rev. Lett.} {\bfseries 126} (2021) 251101} [\href{https://arxiv.org/abs/2103.02616}{{\ttfamily 2103.02616}}].

\bibitem{Ehring:2023abs}
J.~Ehring, S.~Abbar, H.-T. Janka, G.~Raffelt and I.~Tamborra, \emph{{Fast Neutrino Flavor Conversions Can Help and Hinder Neutrino-Driven Explosions}}, \href{https://doi.org/10.1103/PhysRevLett.131.061401}{\emph{Phys. Rev. Lett.} {\bfseries 131} (2023) 061401} [\href{https://arxiv.org/abs/2305.11207}{{\ttfamily 2305.11207}}].

\bibitem{Ehring:2024mjx}
J.~Ehring, S.~Abbar, H.~T. Janka, G.~Raffelt, K.~Nakamura and K.~Kotake, \emph{{Gravitational-Wave Signatures of Nonstandard Neutrino Properties in Collapsing Stellar Cores}},  \href{https://arxiv.org/abs/2412.02750}{{\ttfamily 2412.02750}}.

\bibitem{Fiorillo:2024bzm}
D.~F.~G. Fiorillo and G.~G. Raffelt, \emph{{Theory of neutrino fast flavor evolution. Part I. Linear response theory and stability conditions.}}, \href{https://doi.org/10.1007/JHEP08(2024)225}{\emph{JHEP} {\bfseries 08} (2024) 225} [\href{https://arxiv.org/abs/2406.06708}{{\ttfamily 2406.06708}}].

\bibitem{Fiorillo:2024uki}
D.~F.~G. Fiorillo and G.~G. Raffelt, \emph{{Theory of neutrino fast flavor evolution. Part II. Solutions at the edge of instability}}, \href{https://doi.org/10.1007/JHEP12(2024)205}{\emph{JHEP} {\bfseries 12} (2024) 205} [\href{https://arxiv.org/abs/2409.17232}{{\ttfamily 2409.17232}}].

\bibitem{Fiorillo:2024pns}
D.~F.~G. Fiorillo and G.~G. Raffelt, \emph{{Theory of neutrino slow flavor evolution. Part I. Homogeneous medium}}, \href{https://doi.org/10.1007/JHEP04(2025)146}{\emph{JHEP} {\bfseries 04} (2025) 146} [\href{https://arxiv.org/abs/2412.02747}{{\ttfamily 2412.02747}}].

\bibitem{Fiorillo:2025ank}
D.~F.~G. Fiorillo and G.~G. Raffelt, \emph{{Theory of neutrino slow flavor evolution. Part II. Space-time evolution of linear instabilities}},  \href{https://arxiv.org/abs/2501.16423}{{\ttfamily 2501.16423}}.

\bibitem{Dolgov:1980cq}
A.~D. Dolgov, \emph{{Neutrinos in the early universe}}, {\emph{Sov. J. Nucl. Phys.} {\bfseries 33} (1981) 700}. [{\em Yad.\ Fiz.} {\bf 33} (1981) 1309].

\bibitem{Rudsky}
M.~A. {Rudzsky}, \emph{{Kinetic equations for neutrino spin- and type-oscillations in a medium}}, \href{https://doi.org/10.1007/BF00653658}{\emph{Astrophys. Space Sci} {\bfseries 165} (1990) 65}.

\bibitem{Sigl:1993ctk}
G.~Sigl and G.~Raffelt, \emph{{General kinetic description of relativistic mixed neutrinos}}, \href{https://doi.org/10.1016/0550-3213(93)90175-O}{\emph{Nucl. Phys. B} {\bfseries 406} (1993) 423}.

\bibitem{Sirera:1998ia}
M.~Sirera and A.~P{\'e}rez, \emph{{Relativistic Wigner function approach to neutrino propagation in matter}}, \href{https://doi.org/10.1103/PhysRevD.59.125011}{\emph{Phys. Rev. D} {\bfseries 59} (1999) 125011} [\href{https://arxiv.org/abs/hep-ph/9810347}{{\ttfamily hep-ph/9810347}}].

\bibitem{Yamada:2000za}
S.~Yamada, \emph{{Boltzmann equations for neutrinos with flavor mixings}}, \href{https://doi.org/10.1103/PhysRevD.62.093026}{\emph{Phys. Rev. D} {\bfseries 62} (2000) 093026} [\href{https://arxiv.org/abs/astro-ph/0002502}{{\ttfamily astro-ph/0002502}}].

\bibitem{Vlasenko:2013fja}
A.~Vlasenko, G.~M. Fuller and V.~Cirigliano, \emph{{Neutrino Quantum Kinetics}}, \href{https://doi.org/10.1103/PhysRevD.89.105004}{\emph{Phys. Rev. D} {\bfseries 89} (2014) 105004} [\href{https://arxiv.org/abs/1309.2628}{{\ttfamily 1309.2628}}].

\bibitem{Volpe:2013uxl}
C.~Volpe, D.~V\"a\"an\"anen and C.~Espinoza, \emph{{Extended evolution equations for neutrino propagation in astrophysical and cosmological environments}}, \href{https://doi.org/10.1103/PhysRevD.87.113010}{\emph{Phys. Rev. D} {\bfseries 87} (2013) 113010} [\href{https://arxiv.org/abs/1302.2374}{{\ttfamily 1302.2374}}].

\bibitem{Serreau:2014cfa}
J.~Serreau and C.~Volpe, \emph{{Neutrino-antineutrino correlations in dense anisotropic media}}, \href{https://doi.org/10.1103/PhysRevD.90.125040}{\emph{Phys. Rev. D} {\bfseries 90} (2014) 125040} [\href{https://arxiv.org/abs/1409.3591}{{\ttfamily 1409.3591}}].

\bibitem{Kartavtsev:2015eva}
A.~Kartavtsev, G.~Raffelt and H.~Vogel, \emph{{Neutrino propagation in media: Flavor, helicity, and pair correlations}}, \href{https://doi.org/10.1103/PhysRevD.91.125020}{\emph{Phys. Rev. D} {\bfseries 91} (2015) 125020} [\href{https://arxiv.org/abs/1504.03230}{{\ttfamily 1504.03230}}].

\bibitem{Fiorillo:2024fnl}
D.~F.~G. Fiorillo, G.~G. Raffelt and G.~Sigl, \emph{{Inhomogeneous Kinetic Equation for Mixed Neutrinos: Tracing the Missing Energy}}, \href{https://doi.org/10.1103/PhysRevLett.133.021002}{\emph{Phys. Rev. Lett.} {\bfseries 133} (2024) 021002} [\href{https://arxiv.org/abs/2401.05278}{{\ttfamily 2401.05278}}].

\bibitem{Fiorillo:2024wej}
D.~F.~G. Fiorillo, G.~G. Raffelt and G.~Sigl, \emph{{Collective neutrino-antineutrino oscillations in dense neutrino environments?}}, \href{https://doi.org/10.1103/PhysRevD.109.043031}{\emph{Phys. Rev. D} {\bfseries 109} (2024) 043031} [\href{https://arxiv.org/abs/2401.02478}{{\ttfamily 2401.02478}}].

\bibitem{landau2013electrodynamics}
L.~D. Landau, J.~S. Bell, M.~Kearsley, L.~Pitaevskii, E.~Lifshitz and J.~Sykes, \emph{Electrodynamics of continuous media}, vol.~8. Elsevier, 2013.

\bibitem{supplementalmaterial}
See Supplemental Material for a formal derivation of the flavomon properties, as well as a review of the analogous results for plasmons in a collisionless electron plasma. We also determine the flavomon properties for an example distribution which has a fast instability, and we discuss the properties of flavomons as excitations in a medium in thermal and chemical equilibrium. It includes Refs.~\cite{Johns:2024bob,Johns:2023jjt, Johns:2024dbe, schKT}.

\bibitem{Sawyer:2022ugt}
R.~F. Sawyer, \emph{{Neutrino-anti-neutrino instability in dense neutrino systems, with applications to the early universe and to supernovae}},  \href{https://arxiv.org/abs/2206.09290}{{\ttfamily 2206.09290}}.

\bibitem{Shalgar:2022lvv}
S.~Shalgar and I.~Tamborra, \emph{{Neutrino flavor conversion, advection, and collisions: Toward the full solution}}, \href{https://doi.org/10.1103/PhysRevD.107.063025}{\emph{Phys. Rev. D} {\bfseries 107} (2023) 063025} [\href{https://arxiv.org/abs/2207.04058}{{\ttfamily 2207.04058}}].

\bibitem{Shalgar:2022rjj}
S.~Shalgar and I.~Tamborra, \emph{{Neutrino decoupling is altered by flavor conversion}}, \href{https://doi.org/10.1103/PhysRevD.108.043006}{\emph{Phys. Rev. D} {\bfseries 108} (2023) 043006} [\href{https://arxiv.org/abs/2206.00676}{{\ttfamily 2206.00676}}].

\bibitem{Cornelius:2023eop}
M.~Cornelius, S.~Shalgar and I.~Tamborra, \emph{{Perturbing fast neutrino flavor conversion}}, \href{https://doi.org/10.1088/1475-7516/2024/02/038}{\emph{JCAP} {\bfseries 02} (2024) 038} [\href{https://arxiv.org/abs/2312.03839}{{\ttfamily 2312.03839}}].

\bibitem{Cornelius:2024zsb}
M.~Cornelius, S.~Shalgar and I.~Tamborra, \emph{{Neutrino quantum kinetics in two spatial dimensions}}, \href{https://doi.org/10.1088/1475-7516/2024/11/060}{\emph{JCAP} {\bfseries 11} (2024) 060} [\href{https://arxiv.org/abs/2407.04769}{{\ttfamily 2407.04769}}].

\bibitem{Morinaga:2021vmc}
T.~Morinaga, \emph{{Fast neutrino flavor instability and neutrino flavor lepton number crossings}}, \href{https://doi.org/10.1103/PhysRevD.105.L101301}{\emph{Phys. Rev. D} {\bfseries 105} (2022) L101301} [\href{https://arxiv.org/abs/2103.15267}{{\ttfamily 2103.15267}}].

\bibitem{Dasgupta:2021gfs}
B.~Dasgupta, \emph{{Collective Neutrino Flavor Instability Requires a Crossing}}, \href{https://doi.org/10.1103/PhysRevLett.128.081102}{\emph{Phys. Rev. Lett.} {\bfseries 128} (2022) 081102} [\href{https://arxiv.org/abs/2110.00192}{{\ttfamily 2110.00192}}].

\bibitem{Nagakura:2022kic}
H.~Nagakura and M.~Zaizen, \emph{{Time-Dependent and Quasisteady Features of Fast Neutrino-Flavor Conversion}}, \href{https://doi.org/10.1103/PhysRevLett.129.261101}{\emph{Phys. Rev. Lett.} {\bfseries 129} (2022) 261101} [\href{https://arxiv.org/abs/2206.04097}{{\ttfamily 2206.04097}}].

\bibitem{Zaizen:2022cik}
M.~Zaizen and H.~Nagakura, \emph{{Simple method for determining asymptotic states of fast neutrino-flavor conversion}}, \href{https://doi.org/10.1103/PhysRevD.107.103022}{\emph{Phys. Rev. D} {\bfseries 107} (2023) 103022} [\href{https://arxiv.org/abs/2211.09343}{{\ttfamily 2211.09343}}].

\bibitem{Fiorillo:2024qbl}
D.~F.~G. Fiorillo and G.~G. Raffelt, \emph{{Fast Flavor Conversions at the Edge of Instability in a Two-Beam Model}}, \href{https://doi.org/10.1103/PhysRevLett.133.221004}{\emph{Phys. Rev. Lett.} {\bfseries 133} (2024) 221004} [\href{https://arxiv.org/abs/2403.12189}{{\ttfamily 2403.12189}}].

\bibitem{Padilla-Gay:2021haz}
I.~Padilla-Gay, I.~Tamborra and G.~G. Raffelt, \emph{{Neutrino Flavor Pendulum Reloaded: The Case of Fast Pairwise Conversion}}, \href{https://doi.org/10.1103/PhysRevLett.128.121102}{\emph{Phys. Rev. Lett.} {\bfseries 128} (2022) 121102} [\href{https://arxiv.org/abs/2109.14627}{{\ttfamily 2109.14627}}].

\bibitem{Johns:2019izj}
L.~Johns, H.~Nagakura, G.~M. Fuller and A.~Burrows, \emph{{Neutrino oscillations in supernovae: angular moments and fast instabilities}}, \href{https://doi.org/10.1103/PhysRevD.101.043009}{\emph{Phys. Rev. D} {\bfseries 101} (2020) 043009} [\href{https://arxiv.org/abs/1910.05682}{{\ttfamily 1910.05682}}].

\bibitem{Fiorillo:2023mze}
D.~F.~G. Fiorillo and G.~G. Raffelt, \emph{{Slow and fast collective neutrino oscillations: Invariants and reciprocity}}, \href{https://doi.org/10.1103/PhysRevD.107.043024}{\emph{Phys. Rev. D} {\bfseries 107} (2023) 043024} [\href{https://arxiv.org/abs/2301.09650}{{\ttfamily 2301.09650}}].

\bibitem{Fiorillo:2023hlk}
D.~F.~G. Fiorillo and G.~G. Raffelt, \emph{{Flavor solitons in dense neutrino gases}}, \href{https://doi.org/10.1103/PhysRevD.107.123024}{\emph{Phys. Rev. D} {\bfseries 107} (2023) 123024} [\href{https://arxiv.org/abs/2303.12143}{{\ttfamily 2303.12143}}].

\bibitem{Johns:2021qby}
L.~Johns, \emph{{Collisional Flavor Instabilities of Supernova Neutrinos}}, \href{https://doi.org/10.1103/PhysRevLett.130.191001}{\emph{Phys. Rev. Lett.} {\bfseries 130} (2023) 191001} [\href{https://arxiv.org/abs/2104.11369}{{\ttfamily 2104.11369}}].

\bibitem{Xiong:2022zqz}
Z.~Xiong, L.~Johns, M.-R. Wu and H.~Duan, \emph{{Collisional flavor instability in dense neutrino gases}}, \href{https://doi.org/10.1103/PhysRevD.108.083002}{\emph{Phys. Rev. D} {\bfseries 108} (2023) 083002} [\href{https://arxiv.org/abs/2212.03750}{{\ttfamily 2212.03750}}].

\bibitem{Liu:2023pjw}
J.~Liu, M.~Zaizen and S.~Yamada, \emph{{Systematic study of the resonancelike structure in the collisional flavor instability of neutrinos}}, \href{https://doi.org/10.1103/PhysRevD.107.123011}{\emph{Phys. Rev. D} {\bfseries 107} (2023) 123011} [\href{https://arxiv.org/abs/2302.06263}{{\ttfamily 2302.06263}}].

\bibitem{Lin:2022dek}
Y.-C. Lin and H.~Duan, \emph{{Collision-induced flavor instability in dense neutrino gases with energy-dependent scattering}}, \href{https://doi.org/10.1103/PhysRevD.107.083034}{\emph{Phys. Rev. D} {\bfseries 107} (2023) 083034} [\href{https://arxiv.org/abs/2210.09218}{{\ttfamily 2210.09218}}].

\bibitem{Johns:2022yqy}
L.~Johns and Z.~Xiong, \emph{{Collisional instabilities of neutrinos and their interplay with fast flavor conversion in compact objects}}, \href{https://doi.org/10.1103/PhysRevD.106.103029}{\emph{Phys. Rev. D} {\bfseries 106} (2022) 103029} [\href{https://arxiv.org/abs/2208.11059}{{\ttfamily 2208.11059}}].

\bibitem{Padilla-Gay:2022wck}
I.~Padilla-Gay, I.~Tamborra and G.~G. Raffelt, \emph{{Neutrino fast flavor pendulum. II. Collisional damping}}, \href{https://doi.org/10.1103/PhysRevD.106.103031}{\emph{Phys. Rev. D} {\bfseries 106} (2022) 103031} [\href{https://arxiv.org/abs/2209.11235}{{\ttfamily 2209.11235}}].

\bibitem{Fiorillo:2023ajs}
D.~F.~G. Fiorillo, I.~Padilla-Gay and G.~G. Raffelt, \emph{{Collisions and collective flavor conversion: Integrating out the fast dynamics}}, \href{https://doi.org/10.1103/PhysRevD.109.063021}{\emph{Phys. Rev. D} {\bfseries 109} (2024) 063021} [\href{https://arxiv.org/abs/2312.07612}{{\ttfamily 2312.07612}}].

\bibitem{Johns:2024bob}
L.~Johns, \emph{{Ergodicity demystifies fast neutrino flavor instability}},  \href{https://arxiv.org/abs/2402.08896}{{\ttfamily 2402.08896}}.

\bibitem{Johns:2023jjt}
L.~Johns, \emph{{Thermodynamics of oscillating neutrinos}},  \href{https://arxiv.org/abs/2306.14982}{{\ttfamily 2306.14982}}.

\bibitem{Johns:2024dbe}
L.~Johns, \emph{{Subgrid modeling of neutrino oscillations in astrophysics}},  \href{https://arxiv.org/abs/2401.15247}{{\ttfamily 2401.15247}}.

\bibitem{schKT}
A.~A. {Schekochihin}, \emph{{Lectures on Kinetic Theory and Magnetohydrodynamics of Plasmas}}. Lecture Notes for the Oxford MMathPhys/MScMTP programme; URL: \url{http://www-thphys.physics.ox.ac.uk/people/AlexanderSchekochihin/KT/2015/KTLectureNotes.pdf}, 2024.

\end{thebibliography}\endgroup
